\documentclass[11pt,a4paper]{article}

\usepackage{jheppub}
\usepackage{amsfonts}
\usepackage{amssymb}
\usepackage{comment}
\usepackage{epsfig}
\usepackage{graphicx}
\usepackage{amsmath}
\usepackage{tabu}
\usepackage{subfig}
\usepackage{float}
\usepackage{epstopdf}
\usepackage{multirow}

\newcommand{\bea}{\begin{eqnarray}}
\newcommand{\eea}{\end{eqnarray}}
\newcommand{\bi}{\begin{itemize}}
\newcommand{\ei}{\end{itemize}}
\newcommand{\ben}{\begin{enumerate}}
\newcommand{\een}{\end{enumerate}}
\newcommand{\be}{\begin{equation}}
\newcommand{\ee}{\end{equation}}
\newcommand{\ba}{\begin{align}}
\newcommand{\ea}{\end{align}}
\newcommand{\comments}[1]{}
\def\nn{\nonumber}

\def\FI{{\scriptscriptstyle \rm FI}}

\def\LVS{{\scriptscriptstyle \rm LVS}}

\def\KK{{\scriptscriptstyle \rm KK}}

\def\dS{{\scriptscriptstyle \rm dS}}

\def\I{{\scriptscriptstyle \rm I}}
\def\W{{\scriptscriptstyle \rm W}}
\def\R{{\scriptscriptstyle \rm R}}

\def\L{{\scriptscriptstyle \rm L}}
\def\T{{\mathbb T}}
\def\P{{\mathbb P}}
\def\Z{{\mathbb Z}}

\newcommand\vo{{\mathcal{V}}}
\newcommand{\mbb}{\mathbb}

\newcommand{\mc}{\mathcal}

\newcommand{\beqa}{\begin{eqnarray}}
\newcommand{\eeqa}{\end{eqnarray}}

\title{Global Embedding of Fibre Inflation Models}

\author[a,b,c]{Michele Cicoli,}
\author[d]{Francesco Muia,}
\author[c]{Pramod Shukla}

\affiliation[a]{\small Dipartimento di Fisica e Astronomia, Universit\`a di Bologna, \\ via Irnerio 46, 40126 Bologna, Italy}
\affiliation[b]{\small INFN, Sezione di Bologna, viale Berti Pichat 6/2, 40127 Bologna, Italy}
\affiliation[c]{\small Abdus Salam ICTP, Strada Costiera 11, Trieste 34151, Italy}
\affiliation[d]{\small Rudolf Peierls Centre for Theoretical Physics, University of Oxford, \\ 1 Keble Rd., Oxford OX1 3NP, UK.}

\emailAdd{mcicoli@ictp.it}
\emailAdd{francesco.muia@physics.ox.ac.uk}
\emailAdd{shukla.pramod@ictp.it}

\abstract{We present concrete embeddings of fibre inflation models in globally consistent type IIB Calabi-Yau orientifolds with closed string moduli stabilisation. After performing a systematic search through the existing list of toric Calabi-Yau manifolds, we find several examples that reproduce the minimal setup to embed fibre inflation models. This involves Calabi-Yau manifolds with $h^{1,1}= 3$ which are K3 fibrations over a $\mbb{P}^1$ base with an additional shrinkable rigid divisor. We then provide different consistent choices of the underlying brane set-up which generate a non-perturbative superpotential suitable for moduli stabilisation and string loop corrections with the correct form to drive inflation. For each Calabi-Yau orientifold setting, we also compute the effect of higher derivative contributions and study their influence on the inflationary dynamics.}


\begin{document}

\maketitle

\section{Introduction}
\label{intro}

Inflationary model building is notoriously hard due to the difficulty to protect the flatness of the inflationary direction against potentially large quantum corrections of different origin. This problem is particularly severe for models with observable tensor modes since they generically require the inflaton to travel over a trans-Planckian distance during inflation \cite{Lyth:1996im}. The most promising way-out is based on the presence of symmetries which forbid dangerous corrections to the inflaton potential. 

These symmetries can only be postulated at the effective field theory level but can instead be derived if inflation is embedded within the framework of a consistent UV theory like string theory \cite{McAllister:2007bg, Baumann:2009ni, Cicoli:2011zz, Burgess:2013sla}. From this point of view, the inflaton is the pseudo Nambu-Goldstone boson associated with these symmetries which need to be slightly broken in order to generate the inflaton potential. The small breaking parameter suppresses higher dimensional operators which can spoil the flatness of the inflaton potential. The two main symmetries used for inflationary model building are compact axionic shifts \cite{Westphal:2014ana,Pajer:2013fsa} and non-compact rescaling symmetries for volume moduli \cite{Burgess:2014tja}. 

These symmetries have allowed the realisation of several very promising mechanisms to drive inflation in string compactifications. However some of these mechanisms rely mainly on local constructions which lack a full global realisation in terms of moduli stabilisation. This is crucial to have full control over the inflationary dynamics since it determines the properties of all directions orthogonal to the inflaton and fixes all the mass and energy scales in the model. On top of moduli stabilisation, other important issues to trust inflationary models building are the study of the post-inflationary cosmological history starting with reheating \cite{Green:2007gs, Brandenberger:2008kn, Barnaby:2009wr, Cicoli:2010ha} and the interplay between inflation and other phenomenological implications of the same model like the supersymmetry breaking scale \cite{Conlon:2008cj, He:2010uk, Antusch:2011wu, Buchmuller:2014pla}, the nature of dark matter \cite{Acharya:2008bk, Allahverdi:2013noa, Aparicio:2015sda} and dark radiation \cite{Cicoli:2012aq, Higaki:2012ar, Hebecker:2014gka, Cicoli:2015bpq} or the origin of the matter-antimatter asymmetry \cite{Kane:2011ih, Allahverdi:2016yws}. 

Some string inflation models admit a global realisation with moduli stabilisation but only within the context of an effective supergravity description which assumes the existence of a particular Calabi-Yau background and a suitable form of the superpotential and the K\"ahler potential that define the theory. This is for example the state-of-the-art of fibre inflation models \cite{Burgess:2016owb} where inflation is driven by a K\"ahler modulus and the final prediction for the tensor-to-scalar ratio is between $0.01$ \cite{Cicoli:2016chb} and $0.006$ \cite{Cicoli:2008gp}. In this case, the underlying Calabi-Yau manifolds is assumed to have a fibration structure so that the overall internal volume is controlled by two cycles, the base and the fibre. At leading perturbative order beyond the tree-level approximation, only the overall volume develops a mass while the fibre modulus remains exactly massless. This makes this field a perfect candidate to drive inflation. The perturbative corrections which depend on the fibre modulus and generate its potential are subdominant because of supersymmetry \cite{Cicoli:2007xp}. Moreover the flatness of the fibre modulus potential is protected by an effective shift symmetry associated with the underlying no-scale structure of type IIB compactifications. Even if this symmetry is approximate, since it is broken by loop effects, it is still sufficient to suppress higher dimensional operators \cite{Burgess:2014tja}. 

In this paper we make these inflationary models more robust by embedding them in concrete Calabi-Yau manifolds with an explicit choice of orientifold involution and brane setup which is globally consistent and can, at the same time, reproduce the form of the inflationary potential of fibre inflation models. 
We first derive the topological conditions on the underlying Calabi-Yau manifold which are imposed by the requirement of a successful moduli stabilisation and inflationary mechanism. This singles out Calabi-Yau manifolds with at least $h^{1,1}=3$ that feature a K3 or $T^4$ fibration over a $\mbb{P}^1$ base and a shrinkable rigid (del Pezzo) divisor \cite{Cicoli:2008va, Cicoli:2011it}. We therefore perform a systematic scan through the Kreuzer-Skarke list of toric Calabi-Yau three-folds \cite{Kreuzer:2000xy} to find those with the required structure and find $45$ different examples. We then choose different orientifold involutions and D3/D7 brane setups which satisfy tadpole cancellation conditions and have the right structure to generate the typical potential of fibre inflation models via both string loop \cite{Berg:2005ja, Berg:2007wt, Cicoli:2007xp} and higher derivative $\alpha'$ corrections to the effective action \cite{Ciupke:2015msa}. In the end we perform a detailed analysis of these global models showing that all K\"ahler moduli can be fixed inside the K\"ahler cone and inflation can take place successfully. 

This is the first viable realisation of fibre inflation models in explicit Calabi-Yau orientifold constructions which are globally consistent. This definitely represents an important step forward in our understanding of string inflationary models even if further work in the future is needed. In fact, we shall show that Calabi-Yau examples with $h^{1,1}=3$ are not rich enough to allow non-trivial gauge fluxes on D7-branes which would generate a chiral visible sector. The minimal case which can potentially lead to a global embedding of fibre inflation with a visible chiral sector requires $h^{1,1}=4$. We leave the study of this case for the future. 

This paper is organised as follows. In Sec. \ref{sec2}, after a brief review of the main features of fibre inflation scenarios, we outline the strategy that we shall follow to build viable global models. We then describe the topological and model building requirements of our constructions and present the results of our search through the Kreuzer-Skarke list of toric Calabi-Yau three-folds. We finally explain how we choose the orientifold involution and brane setup and how we compute the resulting string loop corrections to the 4D scalar potential. In Sec. \ref{ExplEx} we then present concrete global models in explicit Calabi-Yau examples with $h^{1,1}=3$. More explicit global examples are described in App. \ref{App}.

\section{Global embedding of fibre inflation}
\label{sec2}

Before presenting our strategy to realise a viable global embedding of fibre inflation models, let us start by reviewing the main features of these inflationary models.

\subsection{A brief review}
\label{Review}

Successful realisations of fibre inflation models require `weak Swiss-cheese' Calabi-Yau (CY) 3-folds whose volume form does not completely diagonalise and in general looks like:
\be
\vo = f_{3/2}\left(\tau_j\right) - \sum_{i=1}^{N_{\rm small}} \lambda_i \tau_i^{3/2}\qquad\text{with}\qquad i\neq j=1,...,h^{1,1} - N_{\rm small}\,,
\label{WSC}
\ee
where $f_{3/2}\left(\tau_j\right)$ is a homogeneous function of degree $3/2$. After fixing at semi-classical level all complex structure moduli and the dilaton by turning on background fluxes \cite{Giddings:2001yu}, the K\"ahler moduli-dependent K\"ahler potential and superpotential are taken of the form:
\be
K = -2\ln \vo + K_{\alpha'} + K_{g_s} \qquad \text{and} \qquad W = W_0 + \sum_{i=1}^{N_{\rm small}} A_i\,e^{-a_i T_i}\,,
\ee
where $\vo$ is the Einstein-frame CY volume in string units, $K_{\alpha'}$ is an $\mc{O}(\alpha'^3)$ correction which depends just on the overall volume $\vo$ \cite{Becker:2002nn, Minasian:2015bxa, Bonetti:2016dqh}, $K_{g_s}$ contains both $\mc{O}(g_s^2 \,\alpha'^2)$ and $\mc{O}(g_s^2 \,\alpha'^4)$ string loop effects which depend on all $T$-moduli \cite{Berg:2005ja, Berg:2007wt, Cicoli:2007xp}, while $W_0$ is the flux superpotential which can be considered as constant after complex structure and dilaton stabilisation. 

According to the general LVS moduli stabilisation procedure, $N_{\rm small}$ blow-up modes plus the overall volume mode get stabilised at leading order giving rise to an AdS vacuum by the interplay of non-perturbative effects in $W$ and $\mc{O}(\alpha'^3)$ corrections to $K$ \cite{Cicoli:2008va}. This leaves $N_{\rm flat} = h^{1,1} - N_{\rm small} - 1$ flat directions which can naturally drive inflation and develop a potential at subleading order by either $\mc{O}(g_s^2 \,\alpha'^4)$ string loop corrections \cite{Berg:2005ja, Berg:2007wt, Cicoli:2007xp} or higher-derivative $F^4\,\mc{O}(\alpha'^3)$ effects \cite{Cicoli:2016chb}.

Note that $\mc{O}(g_s^2 \,\alpha'^2)$ corrections to $K$ contribute effectively to the scalar potential as $\mc{O}(g_s^4 \,\alpha'^4)$ effects since their leading order contribution cancels off because of supersymmetry \cite{Cicoli:2007xp}. This crucial cancellation for inflationary model-building has been name `extended no-scale structure' and can be traced back to the presence of an approximate non-compact shift symmetry \cite{Burgess:2014tja}.

A particularly simple situation arises when $N_{\rm small}=h^{1,1}-2$ since it leads to just $N_{\rm flat} = h^{1,1} - N_{\rm small} - 1 = 1$ flat direction. In this case, the general expression for the CY volume (\ref{WSC}) reduces to:
\be
\vo = \lambda   \sqrt{\tau_1}\,\tau_2 - \sum_{i = 1}^{h^{1,1} - 2} \lambda_i \tau_i^{3/2}\,,
\label{eq:VolumeNflat1}
\ee
where $\tau_1$ is the volume of a $\T^4$ or a K3 fibre over a $\P^1$ base whose volume is given by $t_1=\lambda\tau_2/\sqrt{\tau_1}$ \cite{Math}. Trading the large modulus $\tau_2$ for $\vo \simeq \lambda \sqrt{\tau_1}\, \tau_2$ and working order by order in a large volume expansion, the dominant contribution to the scalar potential at $\mc{O}\left(\vo^{-3}\right)$ can be schematically written as:
\be
V_{\mc{O}\left(\vo^{-3}\right)} = V_\LVS(\vo, \tau_i) + V_\dS(\vo)\,, \quad  \quad i = 1, \dots, N_{\rm small}\,.
\label{eq:PotentialOrder3}
\ee
In the last expression, $V_\LVS(\vo, \tau_i)$ is generated by non-perturbative and $\mc{O}(\alpha'^3)$ effects and gives rise to standard LVS vacua which clearly leave $\tau_1$ unfixed at this level of approximation. The order of magnitude of the LVS potential is \cite{Cicoli:2008va}:
\be
V_\LVS (\vo,\tau_i) \simeq \left(\frac{g_s}{8 \pi}\right) \frac{\xi W_0^2}{g_s^{3/2} \vo^3} \,,
\label{eq:VLVS}
\ee 
where $\xi$ is an $\mc{O}(1)$ topological quantity. $V_\dS$ is instead a model-dependent term which contributes to the vacuum energy and can give rise to a dS solution by properly tuning flux quanta. Its microscopic origin can involve anti-branes \cite{Kachru:2003aw} (for recent progress see \cite{antiDdS}), non-perturbative effects at singularities \cite{Cicoli:2012fh} or T-branes \cite{Cicoli:2015ylx}. 

The flat direction parameterised by $\tau_1$ can drive inflation if it is lifted at subleading order by additional perturbative corrections to $K$ which generate a new contribution $V_{\rm inf}(\tau_1) \ll V_\LVS (\vo,\tau_i)$.\footnote{Non-perturbative corrections to $K$ are negligible in the region where the EFT is under control \cite{Cicoli:2008va}.} The two main effects which can generate $V_{\rm inf}$ are string loop and higher derivative corrections which we briefly discuss below.

\subsubsection{String loop corrections}

Despite the fact that open string 1-loop corrections have been computed explicitly only in simple toroidal cases \cite{Berg:2005ja}, their dependence on K\"ahler moduli for a generic CY manifold has been carefully conjectured in \cite{Berg:2007wt}. In Einstein frame, 1-loop corrections to the K\"ahler potential take two different forms \cite{Berg:2007wt}:
\bea
\text{Kaluza-Klein loops:} \quad K^\KK_{g_s} &=& g_s \sum_i \frac{C_i^\KK t_i^{\perp}}{\vo} \,, 
\label{LoopCorrKkk} \\
\text{Winding loops:} \quad K^\W_{g_s} &=& \sum_i \frac{C_i^\W}{\vo\, t_i^{\cap}}\,.
\label{LoopCorrKw}
\eea
Kaluza-Klein (KK) corrections can be seen in the closed string channel as arising due to the exchange of KK modes between stacks of non-intersecting D3/D7-branes and/or O3/O7-planes. In (\ref{LoopCorrKkk}) $t_i^\perp = a_{ij} t_j$ are the 2-cycles transverse to the stack of parallel D-branes/O-planes. On the other hand, winding corrections can be seen as due to the exchange between stacks of intersecting D-branes/O-planes of closed strings wound around non-contractible 1-cycles at the intersection locus. Accordingly, in (\ref{LoopCorrKw}) $t_i^{\cap} = b_{ij} t_j$ are the 2-cycles where D-branes/O-planes intersect. Moreover, $C_i^\KK$ and $C_i^\W$ are unknown flux-dependent coefficients which can be treated as constants after complex structure and dilaton stabilisation.
 
It is useful to keep track of the order at which these corrections arise both in the $\alpha'$ and $g_s$ expansion. $K_{g_s}^\KK$ is an $\mc{O}\left(g_s^2 \alpha'^2\right)$ effect while $K^\W_{g_s}$ appears at $\mc{O}\left(g_s^2 \alpha'^4\right)$.\footnote{The $\alpha'$ and $g_s$ dependence can be worked out by rewriting the corrections to $K$ in terms of the string frame dimensionful volume ${\rm Vol}_s$ by performing the substitution $\vo \rightarrow {\rm Vol}_s/\left(\alpha'^3 g_s^{3/2}\right)$.} However, the leading KK contribution to the scalar potential vanishes due to the extended no-scale structure, and so the first KK loop correction arises only at $\mc{O}\left(g_s^4 \alpha'^4\right)$ and looks like \cite{Cicoli:2007xp}:
\be
V_{g_s}^\KK = \left(\frac{g_s}{8 \pi}\right) g_s^2 \frac{W_0^2}{\vo^2} \sum_{ij} C_i^\KK C_j^\KK K_{ij} \,,
\label{VgsKK}
\ee 
where $K_{ij}$ is the tree-level K\"ahler metric. Being an $\mc{O}\left(g_s^4 \alpha'^4\right)$ effect, (\ref{VgsKK}) behaves effectively as a 2-loop KK effect. On the other hand, the leading winding contribution to the scalar potential is non-vanishing and reads \cite{Cicoli:2007xp}:
\be
V_{g_s}^\W = -2 \left(\frac{g_s}{8 \pi}\right) \frac{W_0^2}{\vo^2} \, K_{g_s}^\W \,.
\label{VgsW}
\ee

\subsubsection{Higher derivative effects}

The 10D type IIB action for bulk fields receives $\alpha'$ corrections which start contributing at $\mc{O}(\alpha'^3)$ and are encoded in several eight-derivative operators:
\be
S_{\rm IIB}^{10D} = S_0 + \alpha'^3 \, S_3 + \dots\,,
\ee
where the dots indicate the presence of subleading corrections for bulk fields, as well as additional terms related to local sources. $S_3$ denotes a set of eight-derivative operators which can be schematically written as:
\be
S_3\sim \frac{1}{\alpha'^4} \int d^{10}x \, \sqrt{-g} \left[\mc{R}^4 + \mc{R}^3 \left(G_3^2 + ..\right) + \mc{R}^2 \left(G_3^4 + ..\right) + \mc{R} \left(G_3^6 + ..\right) + \left(G_3^8 + ..\right) \right],
\label{8DerivOper}
\ee
where $G_3$ is the type IIB 3-form flux, while the dots in each bracket stand for all possible combinations of fluxes giving rise to an operator with the proper number of derivatives. The second term in (\ref{8DerivOper}), gives rise to the following contribution in the scalar potential \cite{Becker:2002nn}:\footnote{The contribution (\ref{Valpha'}) has been actually derived by first performing a dimensional reduction of the first term in (\ref{8DerivOper}) and then by using supersymmetry arguments.}
\be 
V_{\alpha'} = \left(\frac{g_s}{8 \pi}\right) \frac{3 \xi W_0^2}{4 g_s^{3/2} \vo^3}\,,
\label{Valpha'}
\ee 
which gives rise to LVS minima due to its interplay with non-perturbative effects and fixes the scale of $V_\LVS(\vo,\tau_i)$. This term is of order $F^2$, as can be easily inferred from the $W_0$ dependence. The parameter $\xi$ is completely determined by the CY Euler number $\chi(X)=2(h^{1,1}-h^{1,2})$ since $\xi=-\frac{\chi(X) \zeta(3)}{2(2\pi)^3}$ \cite{Becker:2002nn}. Note that genuinely $N=1$ $\mc{O}(\alpha'^3)$ corrections give rise to an effective Euler number by shifting $\chi(X) \rightarrow \chi_{\rm eff} = \chi(X) + 2 \int_X  D_{\rm O7}^3$, where $D_{\rm O7}$ is the two-form dual to the divisor wrapped by the O7-plane \cite{Minasian:2015bxa}. 

In \cite{Ciupke:2015msa}, the authors were able to infer $F^4$ contributions to the scalar potential which arise from the third term in (\ref{8DerivOper}) and for a general CY take the simple form:\footnote{See also \cite{Weissenbacher:2016gey} for four-derivative terms in the absence of background fluxes. These effects should give rise to small corrections to the moduli canonical normalisation, and so we shall neglect them.}
\be 
V_{F^4} = - \left(\frac{g_s}{8 \pi}\right)^2 \frac{\lambda\,W_0^4}{g_s^{3/2} \vo^4} \sum_{i=1}^{h^{1,1}}\Pi_i t_i \,,
\label{VF4}
\ee 
where $t_i$ are the 2-cycles of the generic CY manifold $X$, while $\Pi_i$ are topological numbers defined as:
\be 
\Pi_i = \int_X c_2 \wedge \hat{D}_i \,.
\label{Pii}
\ee 
Here $c_2$ is the CY second Chern class, $\hat{D}_i$ is a basis of harmonic 2-forms such that the K\"ahler form can be expanded as $J = t_i \hat{D}_i$ and $\lambda$ is an unknown combinatorial factor which is expected to be between $10^{-2}$ and $10^{-3}$ \cite{Ciupke:2015msa}. Note that $\Pi_i \,t_i \geq 0$ in a basis of the K\"ahler cone where $t_i \geq 0$ $\forall i=1,..,h^{1,1}(X)$, implying that all $\Pi_i$ can also be taken as semi-positive definite.

\subsubsection{Inflationary potentials}
\label{InflationaryModels}

Different combinations of perturbative corrections to $K$ can give rise to a different inflationary potential $V_{\rm inf}(\tau_1)$. In fact, depending on compactification details like intersection numbers, divisor topology, brane setup and choice of gauge fluxes and other microscopic parameters, some of the corrections described above can be absent or irrelevant for the stabilisation of $\tau_1$ and the inflationary dynamics. Let us now briefly describe the main features of the different inflationary models proposed so far within this framework.

\subsubsection*{Fibre inflation: KK and winding loops}

The first attempt to realise inflation in fibred CY manifolds with additional shrinkable divisors is `fibre inflation' \cite{Cicoli:2008gp}. In this model the remaining flat direction $\tau_1$ is lifted by the inclusion of both KK and winding loop corrections to $K$. On the other hand, the effect of higher derivative $F^4$ terms is neglected. The scalar potential for the canonically normalised inflaton $\phi$ takes the form:\footnote{Here and in the following $\phi$ represents the displacement of the field from the minimum of the potential.}
\be
V_\FI \simeq \frac{W_0^2}{\langle \vo \rangle^{10/3}} \left[(3 - R) - 4 \left(1 + \frac{R}{6}\right) e^{- k \phi/2} + \left(1 + \frac{2}{3} R\right) e^{- 2 k \phi} + R \,e^{k \phi} \right]\,,
\label{VFI}
\ee
where $k = 2/\sqrt{3}$ and $R$ is a numerical coefficient which is naturally small since $R\propto g_s^4\ll 1$. The minimum of the potential in $\phi=0$ is generated by the competition between the two negative exponentials in (\ref{VFI}) while the term proportional to $e^{- k \phi/2}$ yields an inflationary plateau which can support slow-roll inflation. For large values of $\phi$ the positive exponential in (\ref{VFI}) causes a steepening of the potential which violates the slow-roll conditions. However, due to the smallness of $R$ in the regime with $g_s\ll 1$ where perturbation theory is under control, the inflationary plateau can naturally produce enough efoldings of inflation. The largest tensor-to-scalar ratio which is possible to get with the spectral index compatible with observations is $r \simeq 0.006$ \cite{Cicoli:2008gp}. Note that horizon exit can only occur in the plateau since, if it happened close to the steepening region, the spectral index would become too blue.

\subsubsection*{Left-right inflation: KK loops and higher derivatives}

Ref. \cite{Broy:2015zba} considered the same CY geometry and included $F^4$ higher derivative corrections but neglected winding loops. The resulting inflationary potential contains four terms: two positive KK corrections and two $F^4$ terms whose sign is undetermined. Depending on the sign of these $\alpha'$ effects the potential features an inflationary plateau which can support slow-roll inflation either from left to right or from right to left. Note however that the flatness of the potential tends to be spoiled by dangerous terms which cause a rapid steepening unless one of the two topological quantities controlling $F^4$ terms (see their definition in (\ref{Pii})) is hierarchically smaller than the other by a factor at least of order $10^{-4}$ \cite{Broy:2015zba}. The typical prediction of these generalised fibre inflation models is a relation between the tensor-to-scalar ratio $r$ and the spectral index $n_s$ of the form $r=2 f^2 \left(n_s-1\right)^2$ where $f$ is an effective decay constant controlling the strength of the inflationary plateau generated by the term $e^{-\phi/f}$ \cite{Burgess:2016owb}. Note that in the original fibre inflation model $f=2/k$ \cite{Cicoli:2008gp}. 

The two inflationary potentials proposed in \cite{Broy:2015zba} look like:
\be
V_\R \simeq V_0 \left(1 - e^{k \phi/2}\right)^2 \qquad \text{and}\qquad V_\L \simeq V_0 \left(1 - e^{- k \phi}\right)^2\,,
\label{VLR}
\ee 
where:
\be
V_0 \simeq \left(\frac{\lambda}{g_s}\right)^2 \frac{W_0^6}{\langle\vo\rangle^4} \,.
\ee 
In both cases the minimum is due the interplay between $F^4$ and string loop corrections, while the inflationary plateau is generated by higher derivative effects which take different forms in the two potentials in (\ref{VLR}). In terms of inflationary observables, inflation to the right reproduces the same predictions of fibre inflation since $f=2/k$. On the other hand, inflation to the left has $f=1/k$, and so, from the typical $r$-$n_s$ relation of generalised fibre inflation models, it predicts a tensor-to-scalar ratio smaller by a factor of $4$. Hence for $50$-$60$ e-foldings and a spectral index compatible with observations, the final predictions for tensor modes are: $r_\R \simeq 0.006$ and $r_\L \simeq r_\R/4 \simeq 0.0015$ \cite{Broy:2015zba}.

\subsubsection*{$\alpha'$-inflation: winding loops and higher derivatives}

Another interesting attempt to realise string inflation in fibred CY manifolds with small blow-up modes is `$\alpha'$ inflation' \cite{Cicoli:2016chb}. In this model the inflaton $\tau_1$ develops a potential via higher derivative $F^4$ effects and 1-loop winding corrections. KK loop corrections are neglected since they effectively contribute to the scalar potential only at 2-loop order due to extended no-scale cancellation. The inflationary scalar potential takes the form:
\be
V_{\alpha'\I} \simeq \frac{W_0^2}{\langle \vo \rangle^3 \langle\tau_1\rangle} \left[\left(1 - e^{- k \phi/2}\right)^2  
-R \left(1- e^{k \phi/2} \right)\right]\,,
\label{Vap}
\ee 
where $\langle\tau_1\rangle$ is the value of the inflaton at the minimum and the coefficient $R$ is given by:
\be 
R \simeq \frac{\Pi_2}{\Pi_1} \frac{\langle\tau_1\rangle^{3/2}}{\langle \vo \rangle}\,.
\ee 
If the underlying parameters are chosen to yield an anisotropic compactification at the minimum with $\langle \tau_1\rangle^{3/2}\ll \langle\vo\rangle$, i.e. $\langle \tau_1\rangle\ll \langle\tau_2\rangle$, and the topological quantity $\Pi_2$ is hierarchically smaller than $\Pi_1$, i.e. $\Pi_2\ll \Pi_1$, the coefficient $R$ becomes very small. In the limit $R\to 0$, the potential (\ref{Vap}) represents another example of generalised fibre inflation models with $f=2/k$ where the plateau is generated by winding loops. Hence it reproduces the same predictions of both fibre and left-right inflation, i.e. $r\simeq 0.006$. However, in $\alpha'$ inflation the coefficient of the positive exponential is smaller than in fibre inflation, and so horizon exit can take place also close to the steepening region without obtaining a blue spectral index. In this case the prediction for the tensor-to-scalar ratio can raise to $r \simeq 0.01$ with $n_s \simeq 0.97$ \cite{Cicoli:2016chb}.

\subsection{Global constructions}

After reviewing the main features of fibre inflation scenarios, we are now ready to describe our strategy to build global inflationary models. Let us start by outlining the general topological and model-building requirements.

\subsubsection{General requirements}

In order to realise a successful embedding of fibre inflation models in globally consistent CY orientifolds, we shall follow the following steps: 
\ben
\item Search through the Kreuzer-Skarke (KS) list of toric CY 3-folds \cite{Kreuzer:2000xy} to find those with a fibration structure and at least a shrinkable rigid divisor for LVS moduli stabilisation, i.e. $N_{\rm small}\geq 1$. Requiring in addition at least one flat direction $N_{\rm flat}=h^{1,1}-N_{\rm small}-1\geq 1$, we end up with CY manifolds with Hodge number $h^{1,1}\geq 2+N_{\rm small}\geq 3\,$.

\item Choose an orientifold involution and a D3/D7 brane setup which satisfy both D3- and D7-tadpole cancellation conditions and,\footnote{In the concrete examples of Sec. \ref{ExplEx} we shall not explicitly turn on 3-form background fluxes, and so we will be able to check only the D7-tadpole cancellation condition. However we shall ensure that D3-tadpole cancellation leaves enough space to turn on background fluxes \cite{Cicoli:2011qg}.} at the same time, have enough structure to generate appropriate string loop and higher derivative $\alpha'$ contributions to the scalar potential which can successfully drive inflation.

\item Turn on gauge fluxes on D7-branes to generate a chiral visible sector. 

\item Find a dS vacuum after fixing explicitly all K\"ahler moduli inside the K\"ahler cone.  
\een

If successful, this strategy would lead to the first globally consistent CY orientifold examples with a chiral visible sector and a viable inflationary mechanism together with dS moduli stabilisation. We shall show that CY cases with $h^{1,1}=3$ are not rich enough to satisfy points $(3)$ and $(4)$ above, and so can just lead to a global embedding of fibre inflation models without a chiral visible sector and an explicit dS uplifting mechanism. In order to be able to construct a model where all the points above can potentially be satisfied, one should instead focus on CY cases with at least $h^{1,1}=4$.

\subsubsection{Weak Swiss-cheese CYs}

The simplest scenarios have $N_{\rm flat}=1$ flat direction and $N_{\rm small}=h^{1,1}-2=1$ small blow-up mode which implies $h^{1,1}=3$. These CY examples allow for the realisation of globally consistent inflationary models but are too simple to include a chiral visible sector and an explicit sector responsible for achieving a dS vacuum. In fact, chirality can arise only in the presence of non-vanishing D7 worldvolume fluxes. However these gauge fluxes generate also moduli-dependent Fayet-Iliopoulos terms which together with soft term contributions from matter fields can lift at leading order all K\"ahler moduli charged under anomalous $U(1)$s \cite{Cicoli:2012fh}. This stabilisation method can have the net effect of generating a dS uplifting contribution \cite{CYembedding} corresponding to a T-brane background \cite{Cicoli:2015ylx} but reduces the number of flat directions which can be used to drive inflation. Thus in the $h^{1,1}=3$ case, the only flat direction would become too heavy, and so would not represent anymore a natural inflaton candidate. 

On the other hand, for $N_{\rm small}=h^{1,1}-3$, there are $N_{\rm flat} = h^{1,1}-N_{\rm small}-1=2$ flat directions. Requiring at least $N_{\rm small}=1$ shrinkable divisor, this necessarily leads to CY cases with $h^{1,1}=4$. This situation can now potentially allow for the realisation of inflationary models together with a chiral visible sector and an explicit dS mechanism. In fact, in the presence of chirality, one of the flat directions would be lifted in the process of dS uplifting via D-term driven non-vanishing matter F-terms \cite{CYembedding, Cicoli:2015ylx}, while the other might play the r\^ole of the inflaton. In the $h^{1,1}=3$ case, dS vacua could instead be realised via anti-branes \cite{Kachru:2003aw}. Let us now describe separately these two different situations with $h^{1,1}=3$ and $h^{1,1}=4$ (and $N_{\rm small}=1$) which are both relevant for LVS inflationary constructions.

\subsubsection*{$h^{1,1}=3$ case}

In the $h^{1,1}=3$ case, we shall focus on CY 3-folds $X$ with intersection polynomial of the form:
\be
I_3 = a\, D_f\, D_b^2+\, b\, D_s^3\,,
\label{IntPol}
\ee
where $a$ and $b$ are model dependent integers, while $f$, $b$ and $s$ stay respectively for `fibre', `base' and `small'. This terminology is justified by a theorem by Oguiso which states that if the intersection polynomial is linear in a particular divisor $D_f$, then $D_f$ is either a K3 or a $\T^4$ fibre over a $\P^1$ base \cite{Math}. Moreover (\ref{IntPol}) includes also a shrinkable del Pezzo (dP) 4-cycle $D_s$ suitable to support non-perturbative effects which fix it at small size compared to the overall volume \cite{Cicoli:2008va}. 

Expanding the K\"ahler form $J$ in the divisor basis $\{D_b, \, D_f, \, D_s\}$ as $J = t_f\, D_b+t_b \, D_f+ t_s\, D_s$, the overall volume can be written as:
\be
\vo = \frac{1}{3!}\int_X J \wedge J \wedge J = \frac{a}{2}\, t_f^2\, t_b +\,  \frac{b}{6}\, t_s^3 \,.
\ee
Considering the 4-cycle volume moduli given by:
\bea
\tau_b = \frac{\partial \vo}{\partial t_f} = a\, t_b \, t_f\,, \qquad \tau_f = \frac{\partial \vo}{\partial t_b}=\frac{a}{2}\, t_f^2\,, \qquad \tau_s = \frac{\partial \vo}{\partial t_s}=\frac{b}{2}\,t_s^2\,,
\eea
the CY volume can be rewritten as:
\be
\vo = c_a\,\tau_b\, \, \sqrt{\tau_f} - c_b\,\tau_s^{3/2}\qquad\text{where}\quad c_a= \frac{1}{\sqrt{2\,a}}>0\quad\text{and}\quad c_b = \frac13 
\sqrt{\frac{2}{b}}>0\,.
\ee
The positivity of $\vo$ when $\tau_s\to 0$ forces the integer $a$ to be positive. Moreover, as we shall see below, the fact that $D_s$ is a dP surface 
ensures $b>0$ while its shrinkability implies the K\"ahler cone condition $t_s <0\,$.

\subsubsection*{$h^{1,1}=4$ case}

The intersection polynomial for fibred CY 3-folds $X$ with $h^{1,1} = 4$ and a shrinkable dP surface looks like \cite{Cicoli:2011it}:
\be
I_3 = a\, D_f\, D_{b_1} \, D_{b_2}+\, b\, D_s^3\,.
\ee
Expanding the K\"ahler form $J$ in the divisor basis $\{D_{b_1}, \,D_{b_2},\, D_f, \, D_s\}$ as $J = t_{f_1}\, D_{b_1}+ t_{f_2}\, D_{b_2}+t_b \, D_f+ t_s\, D_s$, the overall volume becomes:
\be
\vo = \frac{1}{3!}\int_X J \wedge J \wedge J = a\, t_{f_1}\, t_{f_2}\, t_b +\,  \frac{b}{6}\, t_s^3 \,.
\ee
The 4-cycle volume moduli read:
\be
\tau_{b_1} = a\, t_{f_2} \, t_b\,, \qquad \tau_{b_2} = a\, t_{f_1} \, t_b\,, \qquad \tau_f = a\, t_{f_1} \, t_{f_2}\,, \qquad \tau_s = \frac{b}{2}\, t_s^2\,,
\ee
and so the CY volume can be rewritten as:
\be
\vo = c_a\,\sqrt{\tau_{b_1}\, \tau_{b_2}\, \tau_f} - c_b\, \tau_s^{3/2}\qquad\text{where}\quad c_a= \frac{1}{\sqrt{a}}>0\quad\text{and}\quad c_b = \frac13\sqrt{\frac{2}{b}}>0\,.
\label{VoLu}
\ee
The relevance of this type of CY volume for cosmological LVS applications has been recently pointed out in \cite{Burgess:2016owb} since (\ref{VoLu}) is analogous to the volume of the simple toroidal example $\T^6/(\Z_2 \times \Z_2)$ ($\vo = \sqrt{\tau_1\, \tau_2\, \tau_3}$) with the only difference being the addition of a blow-up mode. Note that a K3-fibred CY 3-fold with $h^{1,1}=3$ and volume given by (\ref{VoLu}) with $c_b=0$ has been presented in \cite{Gao:2013pra}.

\subsubsection{Divisor topologies}

As we have seen above, `weak Swiss-cheese' CY 3-folds suitable for LVS cosmological applications are K3 or $\T^4$ fibrations over a $\P^1$ base with additional shrinkable dP divisors. Before presenting the results of a scan over the KS list for this kind of CY spaces, let us describe more in depth the topological features of dP, K3 and $\T^4$ surfaces.

\subsubsection*{Del Pezzo divisors}

Del Pezzo divisors are Fano surfaces defined as the algebraic surfaces with ample canonical bundle. These are either dP$_n$ divisors obtained by blowing up $\P^2$ on 8 points ($0 \leq n \leq 8$) or $\P^1 \times \P^1$. Here we shall just consider dP$_n$ surfaces which have the following Hodge diamond:
\bea
{\rm dP}_n \equiv
  \begin{tabular}{ccccc}
    & & 1 & & \\
   & 0 & & 0 & \\
  0 & & $n+1$ & & 0 \\
   & 0 & & 0 & \\
    & & 1 & & \\
  \end{tabular}\qquad \qquad \qquad \forall \, n =0,1,..,8\,. \nonumber
\eea
These surfaces are rigid as $h^{2,0}({\rm dP}_n) = 0$ (in particular dP$_0=\P^2$), and do not contain non-contractible 1-cycles since $h^{1,0}({\rm dP}_n) =0$. Del Pezzo divisors can be of two types: diagonal and non-diagonal \cite{Cicoli:2011it}. Note that a diagonal dP divisor $D_s$ is shrinkable since one can always find a divisor basis where the only term involving $D_s$ in the intersection polynomial is $D_s^3$. 

Divisors with the same Hodge diamond as above but with $n>8$ are still rigid but not del Pezzo. We shall denote them as `NdP$_n$' with $n>8$. These surfaces are not genuine local effects, and so intuitively could be thought of as blow-ups of line-like singularities. Thus NdP$_n$ divisors are generically non-diagonal \cite{Cicoli:2011it}.

A necessary condition for $D_s$ to be dP is that its triple self-intersection is positive and intersections of $(D_i \, D_s^2)$-type with $i\neq s$ are either negative or zero:
\be
\int_{D_i \cap D_s} \, c_1(D_s) = \int_X D_i \wedge D_s \wedge \left(-c_1({\cal N}_{D_s | X})\right) = - \int_X D_i \wedge D_s \wedge D_s \geq 0 \qquad \forall \, i \neq s\,.
\ee 
Here we have used the fact that the first Chern class of a divisor $c_1(D)$ can be written in terms of the first Chern class of its normal bundle $c_1({\cal N}_{D | X})$ as $c_1(D) = - c_1({\cal N}_{D | X}) = - [D]$ where $[D]$ is the homology class of $D$.

Two important quantities characterising the topology of a divisor $D$ are the Euler characteristic $\chi(D)$ and the holomorphic Euler characteristic $\chi_h(D)$ \cite{Blumenhagen:2008zz}:
\bea
\chi(D) &\equiv& \sum_{i=0}^4 {(-1)}^i \, b_i(D)  =  \int_X \, D \wedge\left( D \wedge D + c_2(X) \right)\,, \label{chi} \\
\chi_h({D}) &\equiv& \sum_{i=0}^2 {(-1)}^i \, h^{i,0}(D)  = \frac{1}{12} \int_X \, D \wedge \left(2 \, D\wedge D + c_2(X) \right)\,, 
\label{chih}
\eea
where $b_i(D)$ and $h^{i,0}(D)$ are respectively the Betti and Hodge numbers on the divisor. These two relations also imply:
\be
\chi_h(D) = \frac{1}{12}\left(\chi(D) + \int_X D \wedge D \wedge D\right) \,.
\label{eq:chisAndDDD}
\ee
For connected dP divisors, $\chi_h({\rm dP}_n) = 1$ and $\chi({\rm dP}_n) = n+3$ which from (\ref{eq:chisAndDDD}) give: 
\be
\int_X \, D \wedge D \wedge D = 9-n\,.
\label{chisAndDDDdPn}
\ee
According to (\ref{chisAndDDDdPn}), $D^3_{|_X}>0$ for a dP$_n$ with $n\leq 8$ (for example $D^3_{|_X} = 9$ implies that $D$ is a dP$_0$), while $D^3_{|_X}\leq 0$ indicates that a rigid divisor is a non-shrinkable NdP$_n$ with $n>8$.

\subsubsection*{K3 and $\T^4$ surfaces}

K3 and $\T^4$ surfaces are the only two classes of CY 2-folds. Their Hodge diamonds are:
\bea
{\rm K3} \equiv
\begin{tabular}{ccccc}
    & & 1 & & \\
   & 0 & & 0 & \\
  1 & & 20 & & 1 \\
   & 0 & & 0 & \\
    & & 1 & & \\
  \end{tabular} \qquad \qquad\text{and}\qquad\qquad
		\T^4 \equiv
\begin{tabular}{ccccc}
    & & 1 & & \\
   & 2 & & 2 & \\
  1 & & 4 & & 1 \\
   & 2 & & 2 & \\
    & & 1 & & \\
  \end{tabular}\nonumber
	\eea
As we have seen above, if the CY intersection polynomial is linear in $D_f$, then the CY has the structure of a $D_f$ fibre over a $\P^1$ base \cite{Math}. The condition ${D_f^3}_{|_X}= {D_f^2\,D_i}_{|_X}=0$ $\forall i\neq f$ forces $D_f$ to be either a K3 or a $\T^4$ divisor. In fact, for $D_f={\rm K3}$ we have $\chi(D_f) = 12 \,\chi_h(D_f) = 24$, while for $D_f=\T^4$ we have $\chi(D_f)=\chi_h(D_f)=0$. Thus in both cases (\ref{eq:chisAndDDD}) implies:
\be
\int_X D_f^3 = \chi_h(D_f) - \frac{\chi(D_f)}{12} = 0\,.
\label{eq:fibre1}
\ee
More in general the condition (\ref{eq:fibre1}) is equivalent to require:
\be
\quad h^{1,1}(D_f) = 10\, h^{0, 0}(D_f) - 8\, h^{1,0}(D_f) +10\, h^{2,0}(D_f)\,,
\ee
which is satisfied by K3, $\T^4$ and other topologies. However the additional condition ${D_f^2\,D_i}_{|_X}=0$ $\forall i\neq f$ reduces all possibilities to be either K3 or $\T^4$ \cite{Math}. In this paper we shall present several K3-fibred CY examples, whereas $\T^4$ fibrations are mostly realised in toroidal setups. 

\subsubsection{Scanning results}

Following the topological requirements described above, we considered all the $244$ reflexive lattice polytopes of the KS list \cite{Kreuzer:2000xy} which result in $340$ different CY spaces with $h^{1,1} =3$ after considering the maximal triangulations, and performed a detailed scan to look for `weak Swiss-cheese' CY 3-folds suitable for realising the minimal setup of fibre inflation models. 

We found $102$ CY 3-folds which are K3-fibred and admit at least one dP$_n$ divisor (with $0\leq n\leq 8$). Imposing the further condition that this dP$_n$ divisor should be shrinkable, reduces this number to $45$. We did not perform a proper scan for `weak Swiss-cheese' CY 3-folds with $h^{1,1}=4$ which have the potential to include also a chiral visible sector and an explicit dS uplifting mechanism. We leave this search for future work.

\subsubsection{Orientifold involution and brane setup}

In the explicit examples which we will present in Sec. \ref{ExplEx}, we shall always focus on simple orientifold involutions of the form $\sigma: x_i \to -x_i$ which give rise to an O7-plane wrapped around the `coordinate divisor' $D_i$ defined by $x_i = 0$ plus possible additional O3-planes. We shall then choose an appropriate D7-brane setup which cancels D7-tadpoles and, at the same time, generates string loop corrections to the K\"ahler potential suitable to support fibre inflation models.

In order to cancel all D7-charges, we shall introduce $N_a$ D7-branes wrapped around suitable divisors (say $D_a$) and their orientifold images ($D_a^\prime$) such that \cite{Blumenhagen:2008zz}:
\be
\sum_a\, N_a \left([D_a] + [D_a^\prime] \right) = 8\, [{\rm O7}]\,.
\label{eq:D7tadpole}
\ee
D7-branes and O7-planes also give rise to D3-tadpoles which receive contributions also from background 3-form fluxes $H_3$ and $F_3$, D3-branes and O3-planes. The D3-tadpole cancellation condition reads \cite{Blumenhagen:2008zz}:
\be
N_{\rm D3} + \frac{N_{\rm flux}}{2} + N_{\rm gauge} = \frac{N_{\rm O3}}{4} + \frac{\chi({\rm O7})}{12} + \sum_a\, \frac{N_a \left(\chi(D_a) + \chi(D_a^\prime) \right) }{48}\,,
\label{eq:D3tadpole}
\ee
where $N_{\rm flux} = (2\pi)^{-4} \, (\alpha^\prime)^{-2}\int_X H_3 \wedge F_3$ is the contribution from background fluxes and $N_{\rm gauge} = -\sum_a (8 \pi)^{-2} \int_{D_a}\, {\rm tr}\, {\cal F}_a^2$ is due to D7 worldvolume fluxes. For the simple case where D7-tadpoles are cancelled by placing 4 D7-branes (plus their images) on top of an O7-plane, (\ref{eq:D3tadpole}) reduces to:
\be
N_{\rm D3} + \frac{N_{\rm flux}}{2} + N_{\rm gauge} =\frac{N_{\rm O3}}{4} + \frac{\chi({\rm O7})}{4}\,.
\label{eq:D3tadpole1}
\ee
As a consistency check for a given orientifold involution, one has to ensure that the right-hand-side of (\ref{eq:D3tadpole1}) is an integer. As explained above, in the explicit examples of Sec. \ref{ExplEx} with $h^{1,1}=3$ we shall not turn on gauge fluxes on D7-branes in order to preserve the flatness of the inflaton direction. Thus we shall always have $N_{\rm gauge}=0\,$. Moreover, we shall not explicitly turn on $H_3$ and $F_3$ fluxes but we will always consider orientifold involutions such that the right-hand-side of (\ref{eq:D3tadpole}) is a positive and large integer. This ensures that D3-tadpole cancellation leaves enough freedom to turn on background fluxes for dilaton and complex structure stabilisation. 

\subsubsection{Computation of string loop effects}
\label{ComputeLoops}

Given a particular choice of orientifold involution and brane setup, the location of D-branes and O-planes determines the K\"ahler moduli dependence of open string 1-loop corrections to the scalar potential. In particular, parallel stacks of D-branes/O-planes induce KK corrections, while winding loop effects arise only in the presence of D-branes/O-planes which intersect over a 2-cycle containing non-contractible 1-cycles \cite{Berg:2007wt}. 

In order to understand which involution and brane setup yields a form of these corrections suitable to drive inflation, we shall first compute the intersection curve $D_i\cap D_j$ between each couple of coordinate divisors $D_i$ and $D_j$ and check if this curve can contain non-contractible 1-cycles, i.e. $h^{1,0}(D_i\cap D_j)\neq 0$. If $D_i \cap D_j = \emptyset$ and both divisors are wrapped by D7-branes/O7-planes, string loop corrections are of KK-type and depend on the transverse direction ($t^\perp$) between the two non-intersecting objects. Even if an explicit determination of such a direction requires a detailed knowledge of the CY metric, we are just interested in its dependence on the K\"ahler moduli which for the particular examples of Sec. \ref{ExplEx} can be easily inferred from very general considerations. 

On the other hand, if $D_i \cap D_j \neq \emptyset$, the volume of the intersection 2-cycle is given by:
\be
t^\cap = \int_X J \wedge D_i \wedge D_j\,.
\ee
If both $D_i$ and $D_j$ are wrapped by D7-branes and/or O7-planes, the scalar potential will receive $t^\cap$-dependent winding loop corrections only if $h^{1,0}(D_i\cap D_j)\neq 0$. Finally notice that in the presence of O3-planes, KK corrections always arise due to the exchange of closed KK modes between D7-branes/O7-planes and O3-planes.

\section{Explicit global examples}
\label{ExplEx}

In this section, we shall present all the topological and model-building details of global fibre inflation models in explicit CY orientifolds with $h^{1,1}=3\,$.

\subsection{Toric data}

Let us consider the CY 3-fold $X$ defined by the following toric data:
\begin{table}[H]
  \centering
 \begin{tabular}{|c|ccccccc|}
\hline
     & $x_1$  & $x_2$  & $x_3$  & $x_4$  & $x_5$ & $x_6$  & $x_7$       \\
    \hline
6 & 0  & 0 & 1 & 1 & 1 & 0  & 3   \\
8 & 0  & 1 & 1 & 1 & 0 & 1  & 4   \\
8 & 1  & 0 & 1 & 0 & 1 & 1  & 4   \\   \hline
  & dP$_8$  & NdP$_{10}$ & SD$_1$ &  NdP$_{15}$ & NdP$_{13}$ & K3  &  SD$_2$  \\
    \hline
  \end{tabular}
 \end{table}
 \noindent
with Hodge numbers $(h^{2,1}, h^{1,1}) = (99, 3)$ and Euler number $\chi(X)=-192$. The Stanley-Reisner (SR) ideal is:
\be
{\rm SR} =  \{x_1 x_5,\, x_1 x_6 x_7,\, x_2 x_3 x_4,\, x_2 x_6 x_7,\, x_3 x_4 x_5 \} \,. \nn
\ee
This corresponds to the polytope ID $\#192$ in the CY database of Ref. \cite{Altman:2014bfa}. The intersection polynomial in the basis of smooth divisors $\{D_1, D_6, D_7\}$ is given by:
\be
I_3={D_1^3+ 9 \,D_7^2\, D_1 -3\, D_7\, D_1^2+ 18\, D_7^2\, D_6+  81\, D_7^3}\,,
\label{I3a}
\ee
while the second Chern class is:
\be
c_2 (X) = -\frac{14}{3}\, D_3\, D_7 + \frac23\, D_5\, D_7 +\frac83\, D_7^2\,.
\label{eq:c2a}
\ee
The coordinate divisors are written in terms of the divisor basis as:
\bea
&  & \hskip-0cm D_2 = D_6 - D_1, \qquad D_3 = \frac13 \left(D_7 -  D_6 \right), \quad \\
& &  D_4  = \frac13 \left(D_7 -  D_6  -3\, D_1 \right),   \qquad D_5 =  \frac13 \left(D_7 -4\,  D_6 + 3\, D_1 \right). \nn
\eea

\subsubsection{Coordinate divisors}

A detailed analysis using \texttt{cohomCalg} \cite{Blumenhagen:2010pv, Blumenhagen:2011xn} shows that $D_1$ is a dP$_8$ surface while $D_6$ is a K3. This can also be explicitly seen from the various intersections listed in Tab. \ref{Tab2}.
\begin{table}[H]
  \centering
 \begin{tabular}{|c|ccccccc|}
\hline
     & $D_1$  & $D_2$  & $D_3$  & $D_4$  & $D_5$ & $D_6$  & $D_7$       \\
    \hline
$D_1^2$ & 1  & -1 & -1 & -2 & 0 & 0  & -3   \\
$D_2^2$ & 1  & -1 & -1 & -2 & 0 & 0  & -3   \\
$D_4^2$ & 4  & -2 & -2 & -6 & 0 & 2  & -4   \\
$D_5^2$ & 0  & 2 & -2 & -2 & -4 & 2  & -4   \\
$D_6^2$ & 0  & 0 & 0 & 0 & 0 & 0  & 0   \\
    \hline
  \end{tabular}
  \caption{Intersections between coordinate divisors.}
  \label{Tab2}
 \end{table}
 \noindent
For $D_1$, Tab. \ref{Tab2} shows that $D_1^3= 1 >0$ while $D_1^2\, D_i \leq 0$ $\forall\,i \neq 1$. Thus $D_1$ satisfies the necessary condition to be a dP surface. Moreover, from (\ref{chisAndDDDdPn}) we have that $D_1$ is a dP$_8$ surface with $\chi(D_1)=11$ and $\chi_h(D_1)=1$. On the other hand, the divisor $D_6$ satisfies:
\be
\int_{D_6}\, c_1(D_6) \wedge i^* D_i = -D_6^2\, D_i = 0\,, \qquad \qquad \forall \, i \neq 6\,.
\ee
In addition, using (\ref{chi}) and (\ref{chih}) we find that $\chi(D_6) =24$ and $\chi_h(D_6) = 2$, signaling that $D_6$ is a K3 surface. Furthermore, $D_2$, $D_4$ and $D_5$ are NdP$_n$ divisors which are rigid but not dP surfaces (as introduced in \cite{Cicoli:2011it}). To be more specific, \texttt{cohomCalg} gives the following Hodge numbers:
\bea
 D_2 \equiv
  \begin{tabular}{ccccc}
    & & 1 & & \\
   & 0 & & 0 & \\
  0 & & 11 & & 0 \\
   & 0 & & 0 & \\
    & & 1 & & \\
  \end{tabular} \qquad  D_4 \equiv
  \begin{tabular}{ccccc}
    & & 1 & & \\
   & 0 & & 0 & \\
  0 & & 16 & & 0 \\
   & 0 & & 0 & \\
    & & 1 & & \\
  \end{tabular} \qquad  D_5 \equiv
  \begin{tabular}{ccccc}
    & & 1 & & \\
   & 0 & & 0 & \\
  0 & & 14 & & 0 \\
   & 0 & & 0 & \\
    & & 1 & & \\
  \end{tabular} \nonumber
\eea
Tab. \ref{Tab2} also shows that $D_2$, $D_4$ and $D_5$ do not satisfy the necessary condition for being a dP, although they are rigid. Using triple intersection numbers and the general relations (\ref{chi}) and (\ref{chih}), we find $\chi(D_2) = 13$, $\chi(D_4) = 18$, $\chi(D_5) = 16$ and
$\chi_h(D_2) = \chi_h(D_4)=\chi_h(D_5)=1$. Finally, $D_3$ and $D_7$ are two `special deformation' divisors SD$_1$ and SD$_2$ (in the notation of \cite{Gao:2013pra}) with $h^{01} =0$ and $h^{20} \neq 0$ since their Hodge diamonds read: 
\bea
& & \hskip-0.75cm D_3 \equiv
  \begin{tabular}{ccccc}
    & & 1 & & \\
   & 0 & & 0 & \\
  2 & & 29 & & 2 \\
   & 0 & & 0 & \\
    & & 1 & & \\
  \end{tabular} \qquad \qquad
  D_7 \equiv  \begin{tabular}{ccccc}
    & & 1 & & \\
   & 0 & & 0 & \\
  23 & & 159 & & 23 \\
   & 0 & & 0 & \\
    & & 1 & & \\
  \end{tabular}\,,\nonumber
\eea
showing that $\chi(D_3) = 35$ and $\chi(D_7) = 207$.
 
\subsubsection{Volume form}

Using (\ref{I3a}), and expanding the K\"ahler form in the basis $\{D_1, D_6, D_7\}$ as $J = t_1 \, D_1 + t_6 \, D_6 + t_7 \, D_7$, the overall CY volume becomes:
\be
\vo =\frac{27}{2} \, t_7^3 +9 \,t_7^2 \, t_6+\frac92 \, t_7^2 \, t_1-\frac32 \,t_7 \, t_1^2+\frac16\,t_1^3\,.
\label{vo1}
\ee
Given that the 4-cycle volume moduli look like:
\be
\tau_1 = \frac12\left(t_1 - 3 \,t_7\right)^2\,, \qquad \tau_6 = 9 \, t_7^2\,, \qquad  \tau_7 = \frac{3}{2}\left( 27\, t_7^2 - t_1^2 + 12 t_7 \, t_6 + 6 \, t_7\, t_1\right)\,,
\ee
the overall volume $\vo$ takes the following form:
\be
\vo = \frac16 \left( \sqrt{\tau_6} \, (\tau_7-2 \tau_6 + 3 \tau_1) - 2 \,\sqrt{2} \, \tau_1^{3/2}\right) = t_6\tau_6 +\frac23 \,\tau_6^{3/2} -\frac{\sqrt{2}}{3}\,\tau_1^{3/2}\,.
\label{eq:volEx1a}
\ee
This expression for the volume reflects the fact that $D_6$ is a K3 fibre over a $\P^1$ base of size $t_6$ while $D_1$ is a shrinkable dP$_8$ corresponding to a small divisor in the LVS framework. Moreover it suggests to trade the basis element $D_7$ for $D_x =D_7-2 D_6 + 3 D_1$, since the intersection polynomial (\ref{I3a}) would simplify to:
\be
I_3 = D_1^3 + 18 \, D_6\, D_x^2\,.
\ee
In turn, expanding the K\"ahler form in the new basis as $J = t_s \, D_1 + t_b \, D_6 + t_f \, D_x$, where $s$, $b$ and $f$ stay respectively for `small', `base' and `fibre', the volume form reduces to the minimal version needed for embedding fibre inflation models:
\be
\vo = 9\, t_b \, t_f^2 + \frac16\,t_s^3  = \frac16 \,\sqrt{\tau_f}\tau_b - \frac{\sqrt{2}}{3}\,\tau_s^{3/2} = t_b\tau_f- \frac{\sqrt{2}}{3}\,\tau_s^{3/2}\,,
\label{simpleVol}
\ee
where we have used the following conversion relations in the second step:
\be
t_s = - \sqrt{2}\, \sqrt{\tau_s}\,, \qquad t_b = \frac{\tau_b}{6\, \sqrt{\tau_f}}\,, \qquad  t_f = \frac13\,\sqrt{\tau_f}\,.
\label{ttau}
\ee
Let us finally mention that $D_x$ is a connected divisor with the following Hodge numbers:
\bea
D_x \equiv \begin{tabular}{ccccc}
    & & 1 & & \\
   & 1 & & 1 & \\
  9 & & 92 & & 9 \\
   & 1 & & 1 & \\
    & & 1 & & \\
  \end{tabular}, \qquad \chi(D_x) = 108\,.\nonumber
\eea

\subsection{Brane setups}

Let us now present different globally consistent brane setups which can lead to fibre inflation models. We start by describing possible choices for the orientifold involution.

\subsubsection{Orientifold involution}

We focus on orientifold involutions of the form $x_i \to -x_i$ with $i=1,...,7$ which feature an O7-plane on $D_i$ and O3-planes at the fixed points listed in Tab. \ref{Tab3}. 

\begin{table}[H]
  \centering
 \begin{tabular}{|c|c|c|c|c|c|}
\hline
$\sigma$ & O7 & O3  & $N_{\rm O3}$  & $\chi({\rm O7})$ & $\chi_{\rm eff}$     \\
\hline
\hline
$x_1 \to -x_1$ &  $D_1$ & $\{{D_2 D_3 D_7}, {D_2 D_4 D_5}, {D_3 D_5 D_6}, {D_4 D_6 D_7} \}$ & $\{3,2,2,6\}$ & 11 & -190  \\
$x_2 \to -x_2$ &  $D_2$ & $\{{D_1 D_3 D_7}, {D_3 D_4 D_6}, {D_5 D_6 D_7} \}$ & $\{3,2,6\}$ & 13 & -194  \\
$x_3 \to -x_3$ &  $D_3$ & $\{{D_1 D_2 D_7}, {D_2 D_4 D_6}, {D_4 D_5 D_7}\}$ & $\{3,0,2\}$ & 35  & -190 \\
$x_4 \to -x_4$ &  $D_4$ & $\{{D_2 D_3 D_6}, {D_3 D_5 D_7}\}$ & $\{0,2\}$ & 18 & -204  \\
$x_5 \to -x_5$ &  $D_5$ & $\{{D_1 D_2 D_4}, {D_1 D_3 D_6}, {D_3 D_4 D_7}\}$ & $\{2,0,2\}$ & 16 & -200 \\
$x_6 \to -x_6$ &  $D_6$ & $\{{D_1 D_4 D_7}, {D_2 D_5 D_7}\}$ & $\{6,6\}$ & 24 & -192 \\
$x_7 \to -x_7$ &  $D_7$ & $\{{D_1 D_2 D_3}, {D_1 D_4 D_6}, {D_2 D_5 D_6}\}$ & $\{1,0,0\}$ & 207 & -30 \\
\hline
\end{tabular}
\caption{Fixed point set for involutions of the form $x_i \to -x_i$ with $i=1,...,7$.}  
\label{Tab3}
\end{table}

The effective non-trivial fixed point set in Tab. \ref{Tab3} has been obtained after taking care of the SR ideal symmetry. Moreover, the total number of O3-planes $N_{\rm O3}$ is obtained from the triple intersections restricted to the CY hypersurface, while the effective Euler number $\chi_{\rm eff}$ has been computed, using (\ref{chi}) and (\ref{chih}), as $\chi_{\rm eff} = \chi(X) + 24 \, \chi_h ({\rm O7}) - 2 \, \chi({\rm O7})$. We now focus on two different kinds of D7-brane setups which satisfy the D7-tadpole cancellation condition (\ref{eq:D7tadpole}):
\bi
\item D7-branes on top of the O7-plane: in this case string loop effects simplify since winding corrections are absent due to the fact that there is no intersection between D7-branes and/or O7-planes.

\item D7-branes not (entirely) on top of the O7-plane: in this case $g_s$ corrections to the scalar potential can potentially involve also winding loop effects which are crucial to drive inflation in most fibre inflation models.
\ei 

\subsubsection{Tadpole cancellation}

Let us now present some explicit choices of brane setup which satisfy the D7-tadpole cancellation condition for different orientifold involutions.
\bi
\item \textbf{Case 1}: we focus on the involution $\sigma: x_3 \to -x_3$ with an O7-plane wrapping $D_3$. D7-tadpole cancellation is satisfied via the following brane setup:
\be
8 [{\rm O7}] = 8 \left( [D_2] + [D_5]\right) \,,
\ee
implying that two stacks of D7-branes wrap $D_2$ and $D_5$. The condition for D3-tadpole cancellation (\ref{eq:D7tadpole}) instead becomes:
\be
N_{\rm D3} + \frac{N_{\rm flux}}{2} + N_{\rm gauge} = \frac{N_{\rm O3}}{4} + \frac{\chi({\rm O7})}{12} + \sum_a\, \frac{N_a \left(\chi(D_a) + \chi(D_a^\prime) \right) }{48} = 9\,, \nn
\ee
which leaves some (even if little) space for turning on both gauge and background fluxes for complex structure and dilaton stabilisation. 

\item \textbf{Case 2}: we consider the involution $\sigma: x_6 \to -x_6$ with an O7-plane on $D_6$. The D7 tadpole condition can be satisfied by placing 4 D7-branes (plus their images) on top of the O7-plane:
\be
8 [{\rm O7}] =8 [D_6]\,.
\ee
Thus D3-tadpole cancellation takes the form: 
\be
N_{\rm D3} + \frac{N_{\rm flux}}{2} + N_{\rm gauge} = 9\,, \nn
\ee
leaving again some freedom to turn on gauge and 3-form fluxes.

\item \textbf{Case 3}: the involution $\sigma: x_7 \to -x_7$, which results in an O7-plane wrapping $D_7$, can give rise to a larger freedom for switching on background fluxes since $\chi(D_7) = 207$. The D7-tadpole cancellation condition can be satisfied by: 
\be
a)\qquad 8 [{\rm O7}] = 8\, \left(3\, [D_3] + [D_6] \right)\,,
\ee
with two stacks of D7-branes wrapping $D_3$ and $D_6$. Hence D3-brane tadpoles can be cancelled if:
\be
N_{\rm D3} + \frac{N_{\rm flux}}{2} + N_{\rm gauge} = 39\,, \nn
\ee
showing that the contribution to the total D3-brane charge from background fluxes can indeed be larger in this case. The involution $\sigma: x_7 \to -x_7$ and D7-tadpole cancellation allow also for many other different choices for the brane setup such as:
\bea
b) \qquad 8 [{\rm O7}] &=& 8\left(3\, [D_2] +3\, [D_5]  + [D_6] \right)
\qquad\qquad\Rightarrow\quad N_{\rm D3} + \frac{N_{\rm flux}}{2} + N_{\rm gauge}  = 36\,, \nn \\
c) \qquad 8 [{\rm O7}] &=& 8\left(2\, [D_2] + [D_3] +2\, [D_5]  + [D_6] \right)
\quad\Rightarrow\quad
N_{\rm D3} + \frac{N_{\rm flux}}{2} + N_{\rm gauge}  = 37\,, \nn \\
d) \qquad 8 [{\rm O7}] &=& 8\left([D_2] + 2\, [D_3] + [D_5]  + [D_6] \right)
\quad\,\,\,\,\Rightarrow\quad N_{\rm D3} + \frac{N_{\rm flux}}{2} + N_{\rm gauge}  = 38\,. \nn
\eea
\ei

\subsection{String loop effects}

Let us now follow the procedure described in Sec. \ref{ComputeLoops} to write down the expression for the string loop corrections to the scalar potential for each brane setup described above. Given that winding loop effects arise due to the exchange of strings wound around non-contractible 1-cycles at the intersection between stacks of D7-branes/O7-planes, we start by listing in Tab. \ref{Tab4} all possible intersections between two coordinate divisors. 

\begin{table}[H]
  \centering
 \begin{tabular}{|c|c|c|c|c|c|c|c|}
\hline
  & $D_1$  & $D_2$  & $D_3$  & $D_4$  & $D_5$ & $D_6$  & $D_7$  \\
    \hline
		\hline
 $D_1$ & $\mc{C}_2$  &  $\T^2$      &  $\T^2$        &  $\mc{C}_2$   &  $\emptyset$  &  $\emptyset$   &  $\mc{C}_4$      \\
 $D_2$ & $\T^2$      &  $\P^1$      &  $\T^2$        &  $2\,\P^1$    &  $\mc{C}_2$   &  $\emptyset$   &  $\mc{C}_4$      \\
 $D_3$ & $\T^2$      &  $\T^2$      &  $\mc{C}_2$    &  $\P^1$       &  $\P^1$       &  $\mc{C}_2$    &  $\mc{C}_{14}$   \\
 $D_4$ & $\mc{C}_2$  &  $2\,\P^1$   &  $\P^1$        &  $6\, \P^1$   &  $\P^1$       &  $\mc{C}_2$    &  $\mc{C}_5$      \\
 $D_5$ & $\emptyset$ &  $\mc{C}_2$  &  $\P^1$        &  $\P^1$       &  $4\,\P^1$    &  $\mc{C}_2$    &  $\mc{C}_5$      \\
 $D_6$ & $\emptyset$ &  $\emptyset$ &  $\mc{C}_2$    &  $\mc{C}_2$   &  $\mc{C}_2$   &  $\emptyset$   &  $\mc{C}_{10}$   \\
 $D_7$ & $\mc{C}_4$  &  $\mc{C}_4$  &  $\mc{C}_{14}$ &  $\mc{C}_5$   &  $\mc{C}_5$   &  $\mc{C}_{10}$ &  $\mc{C}_{82}$   \\
    \hline
  \end{tabular}
  \caption{Intersection curves of two coordinate divisors. Here $\mc{C}_g$ denotes a curve with Hodge numbers $h^{0,0} = 1$ and $h^{1,0} = g$ (hence $\T^2\equiv \mc{C}_1$), while $n\,\P^1$ indicates the disjoint union of $n$ $\P^1$s.} 
\label{Tab4}
\end{table}

Notice that, whenever the intersection is the disjoint union of $\P^1$s, there is no non-contractible 1-cycle, and so winding loop corrections are absent by construction. 

\subsubsection{Case 1}

This brane setup is characterised by two D7-stacks wrapping $D_2$ and $D_5$ and an O7-plane located at $D_3$. From Tab. \ref{Tab4} we see that all relevant intersections are:
\be
D_3 \cap D_5 = \P^1\,, \qquad D_2\cap D_3 = \T^2\,, \qquad D_2 \cap D_5 ={\cal C}_2\,.
\ee
Thus we can have winding corrections only from the intersection of the D7s wrapping $D_2$ with either the D7s on $D_5$ or the O7 on $D_3$ since a $\P^1$ does not contain non-contractible 1-cycles. The volumes of the corresponding intersection curves read:
\be
t^\cap(D_2 \cap D_3) = \int_X J \wedge D_2 \wedge D_3 =  t_s + 6 \, t_f\,,  \quad
t^\cap(D_2 \cap D_5) = \int_X J \wedge D_2 \wedge D_5 = 6 \, t_f\,.  \nn
\ee
Therefore from (\ref{LoopCorrKw}) and (\ref{VgsW}) we have that winding string loop corrections take the form:
\be
V_{g_s}^\W=-2\left(\frac{g_s}{8\pi}\right) \frac{W_0^2}{\vo^3} \left( \frac{C_1^\W}{6 \, t_f}+\frac{C_2^\W}{t_s + 6 \, t_f} \right)
= -\left(\frac{g_s}{8\pi}\right) \frac{W_0^2}{\vo^3} \left( \frac{C_1^\W}{\sqrt{\tau_f}}+\frac{C_2^\W}{\sqrt{\tau_f}-\sqrt{\frac{\tau_s}{2}}} \right)\,.
\label{VW}
\ee
Given that the O7-plane and all D7-branes intersect each other, there are no KK loop corrections induced by parallel O7/D7 stacks. However, as shown in Tab. \ref{Tab3}, the fixed point set includes 3 O3-planes located at $D_1 D_2 D_7$ and 2 O3s at $D_4 D_5 D_7$ on the CY hypersurface. Thus KK $g_s$ effects arise from O7/O3 and D7/O3 combinations which lead to a sum over all basis elements in the general 1-loop KK scalar potential (\ref{VgsKK}). Ignoring terms which depend only on $\vo$ and $\tau_s$ that are fixed at leading order, and neglecting terms which have the same volume scaling as (\ref{VW}) but with additional suppression powers of $g_s^2\ll 1$, we end up with (we focus on the region where $\sqrt{\tau_f}\tau_b\gg \tau_s^{3/2}$):
\be
V_{g_s}^\KK = g_s^2 \left(\frac{g_s}{8\pi}\right) \frac{W_0^2}{\vo^2} \left[\frac{(C_f^\KK)^2}{4\tau_f^2} + \frac{(C_b^\KK)^2 \tau_f}{72 \vo^2} \left(1-6\frac{C_s^\KK}{C_b^\KK} \sqrt{\frac{2\tau_s}{\tau_f}}+ \frac{C_f^\KK}{C_b^\KK} \left(\frac{2 \tau_s}{\tau_f}\right)^{3/2}\right) \right]\,.
\label{VKK}
\ee
Therefore the sum of the two string loop corrections (\ref{VW}) and (\ref{VKK}) to the scalar potential for the brane setup 1 is:
\be
V_{g_s}  = \left[\frac{A}{\tau_f^2}-\frac{1}{\vo\sqrt{\tau_f}}\left(\tilde{B} +\frac{\hat{B}}{1-\sqrt{\frac{\tau_s}{2\tau_f}}}\right)+\frac{\tau_f}{\vo^2}
\left(C- \tilde{C} \sqrt{\frac{\tau_s}{\tau_f}}+ \hat{C} \left(\frac{\tau_s}{\tau_f}\right)^{3/2}\right)\right] \frac{W_0^2}{\vo^2}\,,
\label{VA1}
\ee
where (setting $\kappa \equiv g_s/(8\pi)$):
\bea
A &=& \frac{\kappa}{4} \left(g_s\,C_f^\KK\right)^2 >0 \nn \\
\tilde{B} &=&  \kappa\, C_1^\W \nn \\
\hat{B} &=&  - \kappa\, C_2^\W \nn \\
C &=& \frac{\kappa}{72} \left(g_s\, C_b^\KK\right)^2 >0 \nn \\
\tilde{C} &=&  \frac{\kappa}{6\sqrt{2}}\, g_s^2\,C_s^\KK\,C_b^\KK  \nn \\
\hat{C} &=&   \frac{\kappa}{18\sqrt{2}}\,g_s^2\,C_f^\KK\,C_b^\KK\,.  \nn
\eea
Note that in the region of field space where $\tau_f\gg \tau_s$ the terms in (\ref{VA1}) proportional to $\tilde{B}$, $\tilde{C}$ and $\hat{C}$ are negligible and the loop-generated scalar potential simplifies to:
\be
V_{g_s}  \simeq \left(\frac{A}{\tau_f^2}-\frac{B}{\vo\sqrt{\tau_f}} +\frac{C\,\tau_f}{\vo^2} \right) \frac{W_0^2}{\vo^2}\,,
\label{VA1simpl}
\ee
with $B=\tilde{B}+\hat{B}$. This reproduces exactly the inflationary potential of `fibre inflation' \cite{Cicoli:2008gp}.

\subsubsection{Case 2}

In this case D7-tadpole cancellation is ensured by placing 4 D7-branes (plus their images) on top of the O7-plane which wraps $D_6$. Thus there is no intersection between the O7-plane and the D7-branes, resulting in the absence of winding loop corrections. Moreover, there are no KK loop effects from the O7/D7 system since the distance between the O7 and the D7 stack is zero. However, the fixed point set has 6 O3-planes at $D_1 D_4 D_7$ and other 6 O3-planes at $D_2 D_5 D_7$, and so KK string loops can arise from the exchange of KK modes between the O7 or the D7s on $D_6$ and the O3s. Since the volume of $D_6$ is given by $\tau_f$, the simple expression for the volume (\ref{simpleVol}) suggests that the distance between the O7/D7s and the O3s is parametrised by the base of the fibration $t_b$. Hence, using (\ref{VgsKK}), KK string loop correction to the scalar potential become:
\be
V_{g_s}^\KK = \frac{C\,\tau_f\,W_0^2}{\vo^4} \qquad\text{with}\qquad C = \frac{\kappa}{72}\left(g_s\,C_b^\KK\right)^2 \,.
\label{V2}
\ee

\subsubsection{Case 3}

The brane setup of case ($a$) is characterised by two D7-stacks wrapping $D_3$ and $D_6$ and an O7-plane located at $D_7$. From Tab. \ref{Tab4} we see that all relevant intersections are:
\be
D_3 \cap D_6 = \mc{C}_2\,, \qquad D_3\cap D_7 = \mc{C}_{14}\,, \qquad D_6 \cap D_7 =\mc{C}_{10}\,.
\ee
Thus all intersections give rise to curves which contain non-contractible 1-cycles and whose volume takes the form: 
\be
t^\cap(D_3 \cap D_6) =  6 \, t_f \,, \qquad 
t^\cap(D_3 \cap D_7) =  3 \, ( t_s + 6 \, t_f + 2\,t_b) \,,  \qquad
t^\cap(D_6 \cap D_7) =  18 \, t_f\,.  \nn
\ee
Therefore from (\ref{LoopCorrKw}) and (\ref{VgsW}) we have that winding string loop corrections take the form:
\be
V_{g_s}^\W =-\frac23\left(\frac{g_s}{8\pi}\right) \frac{W_0^2}{\vo^3} \left(\frac{3\,C_1^\W+C_2^\W}{6 \, t_f}+\frac{C_3^\W}{t_s + 6 \, t_f + 2\,t_b}\right). 
\label{VW2}
\ee
Working in the limit $t_b\gg t_i$ with $i=f,s$, the previous expression simplifies to:
\be
V_{g_s}^\W \simeq -\frac13\left(\frac{g_s}{8\pi}\right) \frac{W_0^2}{\vo^3} \left(\frac{3\,C_1^\W+C_2^\W}{\sqrt{\tau_f}}+\frac{C_3^\W\,\tau_f}{\vo} \right) \,.
\label{VW3}
\ee
Given that the O7-plane and all D7-branes intersect each other, there are no KK loop corrections induced by parallel O7/D7 stacks. However, as shown in Tab. \ref{Tab3}, the fixed point set includes 1 O3-plane located at $D_1 D_2 D_3$ on the CY hypersurface. Thus KK $g_s$ effects arise from O7/O3 and D7/O3 combinations which lead to a sum over all basis elements in the general 1-loop KK scalar potential (\ref{VgsKK}). Thus KK loop effects take the same form as in (\ref{VKK}) where however the term proportional to $C_b^\KK$ can be neglected since it is suppressed with respect to the term proportional to $C_3^\W$ in (\ref{VW3}) by $g_s^2\ll 1$. Hence the total $g_s$ potential for the brane setup 3 is:
\be
V_{g_s} = V_{g_s}^\W + V_{g_s}^\KK = \left(\frac{A}{\tau_f^2}-\frac{B}{\vo\sqrt{\tau_f}}+\frac{D\,\tau_f}{\vo^2}\right) \frac{W_0^2}{\vo^2}\,,
\label{Vbad}
\ee
where:
\bea
A &=& \frac{\kappa}{4} \left(g_s\,C_f^\KK\right)^2 >0 \nn \\
B &=&  \frac{\kappa}{3} \left(3\,C_1^\W+ C_2^\W\right) \nn \\
D &=&  - \frac{\kappa}{3}\, C_3^\W\,. \nn
\eea
A similar form of the string loop scalar potential arises also for the cases ($b$), ($c$) and ($d$) of the brane setup 3. All these cases allow for a larger freedom to turn on background fluxes but generate a winding correction that has a linear dependence on $\tau_f$ without additional suppression factors of $g_s^2$ which are typical instead of KK loop effects like the one proportional to $C$ in (\ref{VA1simpl}). The winding correction proportional to $D$ in (\ref{Vbad}) would therefore cause a steepening of the potential which would destroy the flatness of the inflationary plateau unless the coefficient $C_3^\W$ is unnaturally tuned to very small values. Thus the potential (\ref{Vbad}) is not particularly suitable to support enough e-foldings of inflation. This example illustrates the challenges that one can encounter in the attempts to build globally consistent D-brane models which give rise to appropriate string loop effects to drive inflation. 

\subsection{Higher derivative corrections}

Let us now consider $F^4$ corrections to the scalar potential which can be computed independently on the choice of the orientifold involution and brane setup. The relevant topological quantities which control the size of these higher derivative $\alpha'$ effects are the various $\Pi_i$'s defined in (\ref{Pii}). They turn out to be:
\be
\Pi_1 =10 , \quad \Pi_2 = 14 , \quad \Pi_3 =34 , \quad \Pi_4 =24 , \quad \Pi_5 =20 ,\quad  \Pi_6 =24 ,  \quad   \Pi_7 =126 , \quad \Pi_x = 108\,. \nonumber
\ee
Now focusing on $\Pi_1$, $\Pi_6$ and $\Pi_x$, from (\ref{VF4}) the higher derivative scalar potential becomes:
\be
V_{F^4} = - \left(\frac{g_s}{8\pi}\right)^2 \frac{\lambda\, W_0^4}{g_s^{3/2}\vo^4} \left(10 \,t_s + 24 \,t_b + 108\, t_f\right)
\simeq \left(\frac{2}{3\vo\,\tau_f}+\frac{\sqrt{\tau_f}}{\vo^2} \right) \frac{F\,W_0^2}{\vo^2}\,,
\label{F4terms}
\ee
where we have neglected the term independent on the fibre modulus $\tau_f$ and: 
\be
F = -36 \,\lambda \left(\kappa\, W_0\right)^2 g_s^{-3/2}\,.
\ee
Note that the topological quantities $\Pi_x$ and $\Pi_6$ which control the size of the two terms in (\ref{F4terms}) are both of the same order since $\Pi_x/\Pi_6=4.5$. As described in Sec. \ref{InflationaryModels}, left-right inflation models require instead a hierarchy between these two topological quantities at least of order $10^4$ in order to protect the flatness of the inflationary potential. Hence we conclude that this explicit CY example does not satisfy a crucial condition for the realisation of left-right inflation models where the inflationary plateau is developed by $F^4$ terms \cite{Broy:2015zba}. 

\subsection{Inflationary dynamics}

The only case whose scalar potential is rich enough to yield an interesting inflationary dynamics is case 1. In fact, in case 2 the string loop potential (\ref{V2}) contains just a single KK contribution and the two $F^4$ terms in (\ref{F4terms}) do not feature the right hierarchy to realise left-right inflation. Moreover in case 3 the scalar potential (\ref{Vbad}) contains a dangerously large winding loop term which depends linearly on $\tau_f$, and so would very quickly destroy the flatness of the potential unless its coefficient is tuned to unnaturally small values. 

We shall therefore focus on the scalar potential of case 1 and show that it can lead to a viable realisation of both `fibre inflation' and `$\alpha'$ inflation' models. The total scalar potential is given by the sum of the loop induced contribution (\ref{VA1}) and the $F^4$ terms (\ref{F4terms}). In the region of field space where $\tau_f \gg \langle \tau_s \rangle$, this scalar potential simplifies to:
\be
V_{\rm tot} = V_{g_s} + V_{F^4} = \frac{W_0^2}{\vo^2} \left[\frac{A}{\tau_f^2}- \frac{B}{\vo \sqrt{\tau_f}} \left(1-\frac{U}{\sqrt{\tau_f}}\right) + \frac{C \tau_f}{\vo^2}  + \frac{F \sqrt{\tau_f}}{\vo^2}\right]\,.
\label{infpot1}
\ee
where (for $C_\W \equiv C_1^\W - C_2^\W$):
\be
|U| = \frac{2|F|}{3B} = \frac{3}{\pi} \frac{|\lambda| \, W_0^2}{C_\W\sqrt{g_s}} \sim \mc{O}(|\lambda|) \ll 1\,,
\ee
for natural values $C_\W\sim W_0\sim\mc{O}(1)$, $g_s\sim\mc{O}(0.1)$ and $|\lambda|\sim\mc{O}(10^{-3})$ \cite{Ciupke:2015msa}. The term proportional to $U$ is therefore suppressed by both $|U|\ll 1$ and $\tau_f\gg 1$, and so it can be safely neglected (it would just slightly shift the position of the minimum). Hence the final form of the inflationary potential is:
\be
V_{\rm inf} = \frac{A\,W_0^2}{\vo^2} \left(\frac{1}{\tau_f^2}- \frac{C_1}{\vo \sqrt{\tau_f}} + C_2 \frac{\tau_f}{\vo^2} + C_3 \frac{\sqrt{\tau_f}}{\vo^2}
\right),
\label{infpot2}
\ee
where:
\be
C_1 = \frac{B}{A}=\left(\frac{2}{C_f^\KK}\right)^2\frac{C_\W}{g_s^2} \qquad C_2 = \frac{C}{A}=\frac 12 \left(\frac{C_b^\KK}{3 C_f^\KK}\right)^2 \qquad
C_3 = \frac{F}{A}=-\frac{162 \,\lambda}{\pi\,g_s^{5/2}} \left(\frac{W_0}{3 C_f^\KK}\right)^2 \,. \nn
\ee
For $g_s\lesssim\mc{O}(0.1)$ and $|\lambda|\sim\mc{O}(10^{-3})$ and natural $\mc{O}(1)$ values of $W_0$ and the coefficients of the string loop effects, the terms in (\ref{infpot2}) proportional to $C_2$ and $C_3$ are both negligible with respect to the $C_1$-dependent term in the region where $1\ll \tau_f\ll\tau_b$ since:
\be
\frac{C_2\,\tau_f/\vo^2}{C_1/(\vo \sqrt{\tau_f})} = \frac{\left(C_b^\KK\right)^2}{C_\W}\frac{g_s^2}{12} \frac{\tau_f}{\tau_b} \ll 1\,,
\ee
and:
\be
\frac{|C_3|\,\sqrt{\tau_f}/\vo^2}{C_1/(\vo \sqrt{\tau_f})} = \frac{972 \,\lambda}{\pi\,C_\W\,g_s^{1/2}} \left(\frac{W_0}{6}\right)^2  \frac{1}{\sqrt{\tau_f}}\frac{\tau_f}{\tau_b} \ll 1\,.
\ee
Therefore the interplay between the first two terms in (\ref{infpot2}) gives a minimum of the inflationary potential for $1\ll \tau_f\ll\tau_b$ which is located at:  
\be
\langle \tau_f \rangle \simeq \left(\frac{4}{C_1}\right)^{2/3}\vo^{2/3}\sim g_s^{4/3} \,\vo^{2/3}\ll \vo^{2/3}
\qquad\Rightarrow\qquad 1\ll\langle \tau_f \rangle\sim g_s^2 \,\langle \tau_b \rangle \ll \langle \tau_b \rangle\,.
\label{minimum}
\ee
In order to study the inflationary dynamics, it is convenient to work with the canonically normalised inflaton $\phi$ defined as:
\be
\tau_f = e^{k \phi} = \langle \tau_f \rangle \,e^{k \varphi}\qquad\text{with}\qquad k = \frac{2}{\sqrt{3}}\,,
\label{CN}
\ee
where we have shifted the inflaton from its minimum as $\phi = \langle \phi \rangle + \varphi$. The scalar potential (\ref{infpot2}) written in terms of $\varphi$ looks like:
\be
V_{\rm inf} = \frac{A W_0^2}{\langle \tau_f \rangle^2 \vo^2} \left(C_\dS+ e^{-2 k \varphi}- 4 e^{-\frac{k \varphi}{2}} + \mc{R}_1\, e^{k \varphi} + \mc{R}_2\, e^{\frac{k \varphi}{2}}\right)\,,
\label{Vref}
\ee
where we added a constant $C_\dS=3 - \mc{R}_1-\mc{R}_2$ to obtain a Minkowski (or slightly dS) vacuum and:
\be
\mc{R}_1 = \frac{16 C_2}{C_1^2}= \left(\frac{C_f^\KK C_b^\KK}{C_\W}\right)^2 \frac{g_s^4}{18}\ll 1 \,, \nn
\ee
and (without loss of generality we choose $\lambda=-|\lambda|<0$):
\be
\mc{R}_2 = \frac{8 C_3}{C_1^{5/3}}\left(\frac{2}{\vo}\right)^{1/3} = \frac{18\,W_0^2}{\pi} \frac{\left(C_f^\KK\right)^{4/3}}{C_\W^{5/3}}
\frac{|\lambda|\,g_s^{5/6}}{\vo^{1/3}} \ll 1 \,. \nn
\label{Rs}
\ee
Note that for $\mc{R}_2 \ll \mc{R}_1\ll 1$ (\ref{Vref}) reproduces exactly the inflationary potential of `fibre inflation' \cite{Cicoli:2008gp}. For $\mc{R}_1 = 10^{-6} \,(10^{-5})$ and $\mc{R}_2 = 0$ we obtain the same predictions: $n_s \simeq 0.964 \,(0.971)$ and $r \simeq 0.007 \,(0.008)$.\footnote{These values are actually slightly different than those reported in \cite{Cicoli:2008gp} since we are evaluating them at $50$, instead of $60$, e-foldings before the end of inflation.} 

On the other hand, for $\mc{R}_1 \ll \mc{R}_2\ll 1$ the potential (\ref{Vref}) is very similar to the one of `$\alpha'$ inflation' since in both cases the plateau is generated by a winding loop effect and the steepening for large $\varphi$ is due to an $F^4$ term \cite{Cicoli:2016chb}. The only difference is the term responsible for the minimum at $\varphi=0$. In our case it is a KK loop correction whereas in \cite{Cicoli:2016chb} it is another higher derivative $\alpha'$ effect. Due to this small difference, the final predictions for the two main cosmological observables are slightly different. In particular, for the same tensor-to-scalar ratio, in our case the spectral index is a bit larger since when the first derivatives of the two inflationary potentials are the same (and so $\epsilon \propto V'^2$ and $r=16\epsilon$ are the same), the second derivative of (\ref{Vref}) is larger than the one of the pure `$\alpha'$ inflation' potential (and so $\eta\propto V''$ and $n_s=1+2\eta-6\epsilon$ are larger). For illustrative purposes, $\mc{R}_2 = 1.5\cdot 10^{-3}$ and $\mc{R}_1 = 0$ would lead to $n_s \simeq 0.976$ and $r \simeq 0.01$ while `$\alpha'$ inflation' predicts $n_s\simeq 0.972$ for the same $r$ \cite{Cicoli:2016chb}.

\begin{figure}[h!]
\begin{center}
\includegraphics[width=0.70\textwidth, angle=0]{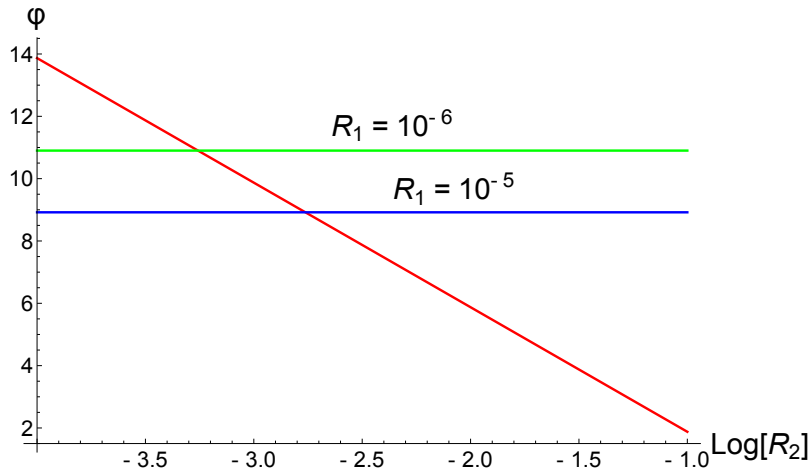}
\caption{The red curve shows the point where the $F^4$ term starts spoiling the flatness of the inflationary plateau as a function of the small parameter $\mc{R}_2$. The green and blue horizontal lines show the values where the KK loop proportional to $\mc{R}_1$ becomes relevant for $\mc{R}_1=10^{-6}$ and $\mc{R}_1=10^{-5}$ respectively.} \label{fig1}
\end{center}
\end{figure}

Let us now consider both positive exponentials in (\ref{Vref}) and compare their relative strength. Note that the term proportional to $\mc{R}_1$ becomes relevant and starts lifting the inflationary plateau roughly when:
\be
0.1 \left(C_\dS - 4 \,e^{-\frac{k \varphi}{2}}\right) \simeq \mc{R}_1\, e^{k \varphi}\,.
\label{loopssteepening}
\ee
Two illustrative numerical results are:
\be
\varphi_1 \simeq 10.9 \quad \text{for} \quad \mc{R}_1 = 10^{-6} \qquad\text{and}\qquad \varphi_2 \simeq 8.9 \quad \text{for} \quad \mc{R}_1 = 10^{-5} \,.
\label{philoops}
\ee
Following the same logic, the point where the positive exponential proportional to $\mc{R}_2$ starts spoiling the flatness of the plateau can be estimated as:
\be
0.1 \left(C_\dS - 4 e^{-\frac{k \varphi}{2}}\right) \simeq \mc{R}_2\, e^{k \varphi/2}\,.
\label{F4steepening}
\ee
Fig. \ref{fig1} shows the value of $\varphi$ where the $F^4$ term becomes relevant as a function of $\mc{R}_2$, and compares it with the two numerical results in (\ref{philoops}). It is clear that the steepening, and hence also the cosmological observables, are determined by the $F^4$ term for $\mc{R}_2 > 5 \cdot 10^{-4}$ if $\mc{R}_1 = 10^{-6}$, and for $\mc{R}_2 > 2 \cdot 10^{-3}$ if $\mc{R}_1 = 10^{-6}$. 

\begin{figure}[h!]
\begin{center}
\includegraphics[width=0.70\textwidth, angle=0]{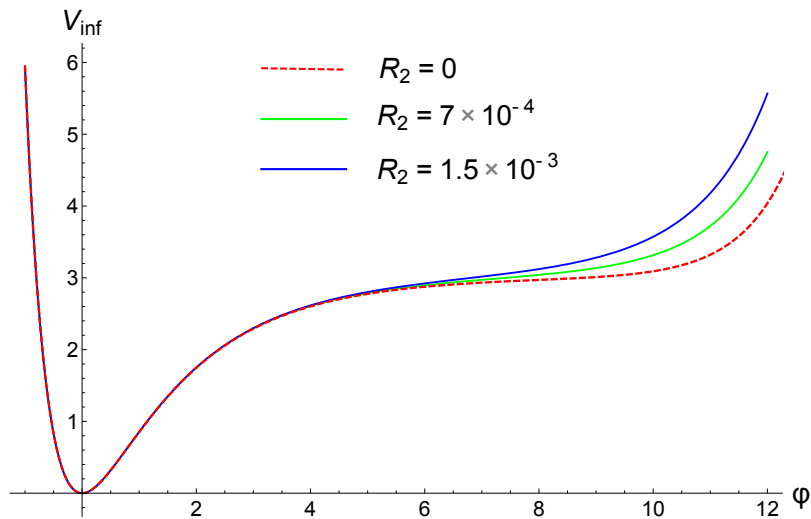}
\caption{Inflationary potential for different values of $\mc{R}_2$ and $\mc{R}_1$ fixed at $\mc{R}_1=10^{-6}$.} \label{fig2}
\end{center}
\end{figure}

In Fig. \ref{fig2} we plot the inflationary potential for the three cases $\mc{R}_2 = \{0, 7 \cdot 10^{-4}, 1.5 \cdot 10^{-3}\}$ and $\mc{R}_1 = 10^{-6}$. The two limiting cases with $\mc{R}_2 = 0$ and $\mc{R}_2=1.5\cdot 10^{-3}$ reproduce respectively `fibre inflation' and `$\alpha'$ inflation'. The corresponding predictions for the cosmological observables are reported in Tab. \ref{tab1}. Note that both $r$ and $n_s$ become larger when $\mc{R}_2$ increases since horizon exit takes place in a steeper region of the scalar potential.

\begin{table}[h!]
\begin{center}
\begin{tabular}{cccccc}
\hline
$\mc{R}_2$ & $n_s$ & $r$ & $\left|W_0\right|$ & $|\lambda|$ & $\delta$ \\
\hline
$0$ & $0.964$ & $0.007$ & $5.7$ & $0$ & $0.17$ \\
\hline
$7 \cdot 10^{-4}$ & $0.970$ & $0.008$ & $6.1$ & $1.5 \cdot 10^{-3}$ & $0.17$ \\
\hline
$1.5 \cdot 10^{-3}$ & $0.977$ & $0.012$ & $6.7$ & $2.7 \cdot 10^{-3}$ & $0.17$ \\
\hline
\end{tabular}
\end{center}
\caption{Predictions for the cosmological observables and choice of the underlying parameters for different values of $\mc{R}_2$ and $\mc{R}_1=10^{-6}$.}
\label{tab1}
\end{table}

The right amplitude of the density perturbations can be obtained by imposing the COBE normalisation:
\be
\left.\frac{V_{\rm inf}^3}{V_{\rm inf}^{'\, 2}}\right|_{\text{horizon exit}} = 2.7 \times 10^{-7}\,,
\label{COBE}
\ee
which, for $g_s = 0.1$ and $\vo = 10^3$, requires natural values of $W_0$ reported in Tab. \ref{tab1}. Moreover $\mc{R}_1 = 10^{-6}$ and $\mc{R}_2$ can be exactly reproduced with reasonable choices of the underlying parameters $C_\W = 90$, $C^\KK_f = 65$, $C^\KK_b = 0.58$ and the values of $|\lambda|$ listed in Tab. \ref{tab1}. 

Let us finally check the consistency of the effective field theory. In order to trust our single field approximation, the mass of the volume mode has to be larger than the Hubble scale $H^2 \simeq V_{\rm inf}/3$. This condition boils down to:
\be
\delta = \frac{H^2}{m^2_{\vo}} \simeq \frac{V_{\rm inf}}{V_{\alpha'}} \ll 1\,,
\label{singlefield}
\ee
where the $V_{\alpha'}$ is the leading $\mc{O}(\alpha'^3)$ contribution to the scalar potential and reads:
\be
V_{\alpha'} = \kappa \frac{3 \xi W_0^2}{4 g_s^{3/2} \vo^3}\,.
\ee
As shown in Tab. \ref{tab1}, the single field approximation is under control since $\delta\ll 1$ for each case (our CY example has $\chi_{\rm eff} = -190$ which gives $\xi = 0.46$). Moreover the $\alpha'$ expansion can be trusted only if:
\be
\zeta = \frac{\xi}{2 g_s^{3/2} \vo} \ll 1\,.
\ee
The previous choice of $g_s$ and $\vo$ gives $\zeta \simeq 0.007$, implying that also the $\alpha'$ expansion is under control. Finally, our choice of microscopic parameters leads to $\langle \tau_f \rangle \simeq 60 \gg \langle \tau_s \rangle \simeq 3$, so that the corrections proportional to $\langle \tau_s \rangle/\tau_f\lesssim 0.05$ in (\ref{VA1}) can be consistently neglected.

\section{Conclusions}
\label{Concl}

String inflation models are very promising to describe the early universe due to the presence of approximate symmetries which can explain the flatness of the inflationary potential. However they sometimes lack a fully consistent description of the mechanism responsible to stabilise all the moduli in a concrete Calabi-Yau compactification. 

In this paper we presented the first explicit type IIB examples of globally consistent models where the Calabi-Yau background is described by toric geometry, all closed string moduli are stabilised, and the scalar potential of one of the K\"ahler moduli, namely the fibre modulus, is suitable to drive inflation. In particular we managed to produce global embeddings of both `fibre inflation' \cite{Cicoli:2008gp} and `$\alpha'$ inflation' \cite{Cicoli:2016chb} models which can predict primordial gravitational waves which might be detectable in the near future. 

After finding $45$ different `weak Swiss cheese' Calabi-Yau manifolds with $h^{1,1}= 3$ with a shrinkable del Pezzo divisor needed for LVS moduli stabilisation, we focused on one of them and provided different consistent models with O7/O3-planes and D3/D7-branes which lead to the generation of the correct perturbative (both in $\alpha'$ and $g_s$) and non-perturbative effects to stabilise all the K\"ahler moduli and reproduce the potential of fibre inflation models. In particular, we showed that the inflationary potential features a long plateau that is naturally generated by winding loop corrections.  At large inflaton values the inflationary potential has instead a raising behaviour which, depending on the values of the underlying parameters, can be due to either KK loop effects of higher derivative contributions. 

We believe that this paper represents an important step forward in the construction of globally consistent string inflation models even if Calabi-Yau manifolds with $h^{1,1}=3$ are too simple to allow also for the realisation of a chiral visible sector. Chiral matter can be included only in the presence of non-vanishing gauge fluxes on D7-branes which, when combined with the requirement of a viable inflationary direction, require Calabi-Yau manifolds with $h^{1,1}=4$. We leave the study of this case for future work.

\section*{Acknowledgements}

We would like to thank Roberto Valandro for many useful discussions. We are also thankful to Ross Altman, Volker Braun, David Ciupke, Xin Gao, James Gray, Jim Halverson and Christoph Mayrhofer for several useful conversations. PS is grateful to the Bologna INFN division for hospitality during the Spring and the Summer of 2016 when most of this work was carried out.

\appendix

\section{More explicit global examples}
\label{App}

In this appendix we provide additional explicit global examples of fibre inflation models.

\subsection{Toric data}

Let us consider the CY 3-fold $X$ defined by the following toric data:
\begin{table}[H]
  \centering
 \begin{tabular}{|c|ccccccc|}
\hline
     & $x_1$  & $x_2$  & $x_3$  & $x_4$  & $x_5$ & $x_6$  & $x_7$       \\
    \hline
 6  & 0  & 0 & 1 & 1 & 0 & 3  & 1   \\
 8  & 0  & 1 & 1 & 1 & 1 & 4  & 0   \\
12 & 1  & 0 & 2 & 2 & 1 & 6  & 0   \\   \hline
    &  NdP$_{11}$  & dP$_7$  & SD$_1$ &  SD$_1$ & K3 & SD$_2$  & W  \\
    \hline
  \end{tabular}
 \end{table}
 \noindent
with Hodge numbers $(h^{2,1}, h^{1,1}) = (111, 3)$ and Euler number $\chi(X)=-216$.
The Stanley-Reisner ideal is:
\be
{\rm SR} =  \{ x_1 x_5, \, x_2 x_5, \, x_1 x_3 x_4, \, x_2 x_6 x_7, \, x_3 x_4 x_6 x_7\} \, \nn
\ee
The intersection polynomial in the basis of smooth divisors $\{D_2, D_5, D_7\}$ is given by:
\be
I_3=2 D_2^3 + 2 D_5\, D_7^2 - 8 D_7^3\,,
\label{eq:I3a}
\ee
while the second Chern class reads:
\be
c_2(X) = 4\, D_4^2 + 34 \, D_5 \, D_7 + 8 D_7^2\,.
\label{eq:c2a}
\ee
The various classes for the coordinate divisor are written in terms of divisor basis as:
\be
D_1 = D_5 - D_2, \quad D_3  = 2 D_5 +D_7- D_2 \equiv D_4, \quad D_6 = 6D_5 + 3 D_7-2 D_4\,. 
\label{Ds}
\ee

\subsubsection{Coordinate divisors}

A detailed analysis using \texttt{cohomCalg} \cite{Blumenhagen:2010pv, Blumenhagen:2011xn} shows that $D_2$ is a dP$_7$ surface while $D_5$ is a K3. This can also be explicitly seen in Tab. \ref{Tab5} below. 
\begin{table}[H]
  \centering
 \begin{tabular}{|c|ccccccc|}
\hline
     & $D_1$  & $D_2$  & $D_3$  & $D_4$  & $D_5$ & $D_6$  & $D_7$       \\
    \hline
$D_1^2$ & -2  & 2 & -2 & -2 & 0 & -4  & 0   \\
$D_2^2$ & -2  & 2 & -2 & -2 & 0 & -4  & 0   \\
$D_5^2$ & 0  & 0 & 0 & 0 & 0 & 0  & 0   \\
    \hline
  \end{tabular}
  \caption{Intersections between K3 and rigid divisors.}
  \label{Tab5}
 \end{table}
 \noindent
Tab. \ref{Tab5} shows that the triple intersection number $D_2^3 = 2 >0$ while any other intersection number of the kind $D_2^2\, D_i$, with $i \neq 2$, is either negative or zero, thus satisfying the necessary condition for $D_2$ to be a dP surface. Moreover, (\ref{chisAndDDDdPn}) ensures that $D_2$ is a dP$_7$ surface. Moreover from (\ref{eq:I3a}) and (\ref{eq:c2a}) we find that the Euler characteristic and the holomorphic Euler characteristic of $D_2$ turn out to be:
\bea
& & \chi(D_2) = \int_X \left(D_2^3 + c_2(X) \wedge D_2 \right) = 2 + 8 =10\,, \\
& &  \chi_h(D_2) = \frac{1}{12} \int_X \left(2 D_2^3 + c_2(X) \wedge D_2 \right) = \frac{4+8}{12}= 1\,. \nn
\eea
Tab. \ref{Tab5} also shows that $D_1$ does not satisfy the necessary condition for being a dP, though it is a rigid divisor as can be cross-checked from the fact that $\chi(D_1) =14$ and $\chi_h(D_1) = 1$. Regarding instead the divisor $D_5$, from Tab. \ref{Tab5} we find:
\be
\int_{D_5}\, c_1(D_5) \wedge i^{*}D_i = -D_5^2\, D_i = 0\,, \qquad \qquad \forall \, i \neq 5\,.
\ee
In addition, $\chi(D_5) = 24$ and $\chi_h(D_5) = 2$, and so $D_5$ is expected to be a K3 surface. Furthermore, the divisor $D_7$ is a so-called Wilson-divisor ($\rm W$) which has no deformation moduli but has $h^{1,0}({\rm W})\neq 0$ \cite{Gao:2013pra, Blumenhagen:2012kz}. To be more specific, \texttt{cohomCalg} gives the following Hodge numbers for the divisor W:
\bea
 {\rm W} \equiv
  \begin{tabular}{ccccc}
    & & 1 & & \\
   & 2 & & 2 & \\
  0 & & 2 & & 0 \\
   & 2 & & 2 & \\
    & & 1 & & \\
  \end{tabular} \nn
\eea
which can be cross-checked by using the triple intersection numbers via the Euler characteristic $\chi(D_7) =-4$ and the holomorphic Euler characteristic $\chi_h(D_7) = -1$. Similarly, the divisors $D_3=D_4$ and $D_6$ denoted as SD$_1$ and SD$_2$ in the toric data table come out to have the following Hodge diamonds:
\bea
\hskip-0.75cm 
  {\rm SD}_1 \equiv
  \begin{tabular}{ccccc}
    & & 1 & & \\
   & 0 & & 0 & \\
  3 & & 38 & & 3 \\
   & 0 & & 0 & \\
    & & 1 & & \\
  \end{tabular} \,, \qquad  \qquad 
{\rm SD}_2 \equiv  \begin{tabular}{ccccc}
    & & 1 & & \\
   & 0 & & 0 & \\
  26 & & 178 & & 26 \\
   & 0 & & 0 & \\
    & & 1 & & \\
  \end{tabular}\,. \nonumber
\eea
which imply $\chi(D_3)=46=\chi(D_4)$ and $\chi(D_6)= 232$. 

\subsubsection{Volume form}

Using the intersection polynomial in (\ref{eq:I3a}) and expanding the K\"ahler form in the divisor basis as $J = t_2 \, D_2 + t_5 \, D_5 + t_7 \, D_7$, the overall CY volume takes the form:
\be
\vo = \frac{t_2^3}{3} + t_5\, t_7^2 - \frac43 \, t_7^3\,.
\label{VoL}
\ee
Trading the 2-cycle for the 4-cycle volumes using:
\be
\tau_2 = t_2^2\,, \qquad \tau_5 = t_7^2\,, \qquad  \tau_7 = 2 \,(t_5 - 2\, t_7) \, t_7\,,  \nn
\ee
the volume form (\ref{VoL}) becomes:
\be
\vo = \frac16 \left( \sqrt{\tau_5} \, (3 \tau_7 + 4 \tau_5) - 2 \, \tau_2^{3/2}\right) = t_5\,\tau_5 - \frac43\, \tau_5^{3/2}-\frac13\, \tau_2^{3/2}\,,
\label{volEx2a}
\ee
reflecting the fact that $D_5$ is a K3 fibre over a $\P^1$ base with volume given by $t_5$, while $D_2$ plays the r\^ole of the small blow-up mode typical of LVS constructions. Moreover (\ref{volEx2a}) suggests to switch to the basis $\{D_2, D_5, D_x =(4\,D_5+3\,D_7) \}$ where the intersection polynomial reduces to:
\be
I_3 = 2 D_2^3 + 18 D_5\, D_x^2\,.
\ee
Expanding the K\"ahler form as $J = t_s \, D_2 + t_b \, D_5 + t_f \, D_x$, the volume form (\ref{volEx2a}) reduces into the minimal version needed for embedding fibre inflation models:
\be
\vo = 9\, t_b \, t_f^2 + \frac13\, t_s^3 = \frac16\, \sqrt{\tau_f}\,\tau_b - \frac13\,\tau_s^{3/2}\,,
\label{SimpVol}
\ee
where the relation between 2- and 4-cycles becomes:
\be
t_s = - \sqrt{\tau_s}\,, \qquad t_b =\frac{\tau_b}{6 \sqrt{\tau_f}}\,, \qquad t_f = \frac13\,\sqrt{\tau_f}\,.
\ee

\subsection{Brane setups}

Let us now present different globally consistent brane setups which can lead to fibre inflation models. We start by describing possible choices for the orientifold involution.

\subsubsection{Orientifold involution}

We focus on orientifold involutions of the form $x_i \to -x_i$ with $i=1,...,7$ which feature an O7-plane on $D_i$ and O3-planes at the fixed points listed in Tab. \ref{Tab6}. 

\begin{table}[H]
  \centering
 \begin{tabular}{|c|c|c|c|c|c|}
\hline
$\sigma$ & O7  & O3  & $N_{\rm O3}$  & $\chi({\rm O7})$ & $\chi_{\rm eff}$      \\
\hline
\hline
$x_1 \to -x_1$ &  $D_1 \sqcup D_5$ &  $D_2 D_3 D_4$ & 2 & 24+14 & -220 \\
$x_2 \to -x_2$ &  $D_2$ & $\{D_1 D_6 D_7, D_3 D_4 D_5, D_5 D_6 D_7\}$ & \{6,2,6\} & 10 & -212 \\
$x_3 \to -x_3$ &  $D_3$ & $\{D_1 D_2 D_4, D_4 D_6 D_7\}$ & \{2,0\} & 46 & -212 \\
$x_4 \to -x_4$ &  $D_4$ & $\{D_1 D_2 D_3, D_3 D_6 D_7\}$ & \{2,0\} & 46 & -212 \\
$x_5 \to -x_5$ &  $D_5 \sqcup D_1$ &  $D_2 D_3 D_4$ & 2 & 14+24 & -220 \\
$x_6 \to -x_6$ &  $D_6$ & $\{D_1 D_2 D_7, D_3 D_4 D_7\}$ & \{0,0\} & 232 & -32 \\
$x_7 \to -x_7$ &  $D_7$ &  $\{D_1 D_2 D_6, D_3 D_4 D_6\}$ & \{4,8\} & -4 & -232 \\
\hline
\end{tabular}
\caption{Fixed point set for involutions of the form $x_i \to -x_i$ with $i=1,...,7$.}  
\label{Tab6}
\end{table}

The effective non-trivial fixed point set in Tab. \ref{Tab6} has been obtained after taking care of the SR ideal symmetry. Moreover, the total number of O3-planes $N_{\rm O3}$ is obtained from the triple intersections restricted to the CY hypersurface. We now focus on two different kinds of D7-brane setups which satisfy the D7-tadpole cancellation condition (\ref{eq:D7tadpole}):
\bi
\item D7-branes on top of the O7-plane: in this case string loop effects simplify since winding corrections are absent due to the fact that there is no intersection between D7-branes and/or O7-planes.

\item D7-branes not (entirely) on top of the O7-plane: in this case $g_s$ corrections to the scalar potential can potentially involve also winding loop effects which are crucial to drive inflation in most fibre inflation models.
\ei 

\subsubsection{Tadpole cancellation}

Let us now present some explicit choices of brane setup which satisfy the D7-tadpole cancellation condition for different orientifold involutions.
\bi
\item \textbf{Case 1}: we focus on the involution $\sigma: x_4 \to -x_4$ (or equivalently $\sigma: x_3 \to -x_3$) with an O7-plane wrapping $D_4$. D7-tadpole cancellation is satisfied via the following brane setup:
\be
8 [{\rm O7}] = 8 \left( [D_1] + [D_5] + [D_7]\right) \,,
\ee
implying that three stacks of D7-branes wrap $D_1$, $D_5$ and $D_7$. The condition for D3-tadpole cancellation (\ref{eq:D7tadpole}) instead becomes:
\be
N_{\rm D3} + \frac{N_{\rm flux}}{2} + N_{\rm gauge} = \frac{N_{\rm O3}}{4} + \frac{\chi({\rm O7})}{12} + \sum_a\, \frac{N_a \left(\chi(D_a) + \chi(D_a^\prime) \right) }{48} = 10\,, \nn
\ee
which leaves some (even if little) space for turning on both gauge and background fluxes for complex structure and dilaton stabilisation. 

\item \textbf{Case 2}: as can be seen from Tab. \ref{Tab6}, the involution $\sigma: x_5 \to -x_5$ leads to two O7-planes at $D_1$ and $D_5$. The D7-tadpole cancellation condition can be satisfied by placing 4 D7-branes (plus their images) on top of each O7-plane:
\be
8 [{\rm O7}] =  8 \left( [D_1]+ [D_5] \right) \,,
\ee
which leads to the following D3-tadpole cancellation condition:
\be
N_{\rm D3} + \frac{N_{\rm flux}}{2} + N_{\rm gauge} = 10\,. \nn
\ee

\item \textbf{Case 3}: the involution $\sigma: x_6 \to -x_6$ gives an O7-plane at $D_6$ without any O3-plane as the relevant fixed points $\{ {D_1 D_2 D_7}, \, D_3 D_4 D_7 \}$ do not intersect the CY hypersurface. The D7-tadpole cancellation condition can be ensured by considering the following brane setup:
\be
8 [{\rm O7}] = 8 \left(2\, [D_1] + 4\, [D_5] + 3\, [D_7] \right)\,,
\ee
with three stacks of D7s wrapping $D_1$, $D_5$ and $D_7$. D3-brane tadpole cancellation leave a lot of freedom to turn on background fluxes since it reads: 
\be
N_{\rm D3} + \frac{N_{\rm flux}}{2} + N_{\rm gauge} =  38\,. \nn
\ee
\ei

\subsection{String loop effects}

Let us now follow the procedure described in Sec. \ref{Review} to write down the expression for the string loop corrections to the scalar potential for each brane setup described above. Given that winding loop effects arise due to the exchange of strings wound around non-contractible 1-cycles at the intersection between stacks of D7-branes/O7-planes, we start by listing in Tab. \ref{Tab7} all possible intersections between two coordinate divisors. 

\begin{table}[H]
  \centering
 \begin{tabular}{|c|c|c|c|c|c|c|c|}
 \hline
  & $D_1$  & $D_2$  & $D_3$  & $D_4$  & $D_5$ & $D_6$  & $D_7$       \\
 \hline
 \hline
 $D_1$ & $2\,\P^1$   & $\T^2$        & $\P^1$          & $\P^1$          & $\emptyset$     & $\mc{C}_4$       & $\mc{C}_2$  \\
 $D_2$ & $\T^2$      & $\mc{C}_3$    & $\T^2$          & $\T^2$          & $\emptyset$     & $\mc{C}_3$       & $\emptyset$   \\
 $D_3$ & $\P^1$      & $\T^2$        & $\mc{C}_3$      & $\mc{C}_3$      & $\mc{C}_2$      & $\mc{C}_{19}$    & $2\, \P^1$  \\
 $D_4$ & $\P^1$      & $\T^2$        & $\mc{C}_3$      & $\mc{C}_3$      & $\mc{C}_2$      & $\mc{C}_{19}$    & $2\, \P^1$  \\
 $D_5$ & $\emptyset$ & $\emptyset$   & $\mc{C}_2$      & $\mc{C}_2$      & $\emptyset$     & $\mc{C}_{10}$    & $\mc{C}_2$  \\
 $D_6$ & $\mc{C}_4$  & $\mc{C}_3$    & $\mc{C}_{19}$   & $\mc{C}_{19}$   & $\mc{C}_{10}$   & $\mc{C}_{93}$    & $6\,\P^1$  \\
 $D_7$ & $\mc{C}_2$  & $\emptyset$   & $2\,\P^1$       & $2\,\P^1$       & $\mc{C}_2$      & $6\,\P^1$        & $8\,\P^1$  \\
\hline
\end{tabular}
\caption{Intersection curves of two coordinate divisors. Here ${\cal C}_g$ denotes a curve with the Hodge numbers $h^{00} = 1$ and $h^{10} = g$ while $n\,\P^1$ denotes the disjoint union of $n$ $\P^1$s.}
\label{Tab7}
\end{table}

\subsubsection{Case 1}

This brane setup is characterised by 3 D7-stacks wrapping $D_1$, $D_5$ and $D_7$ and an O7-plane located at $D_4$. From Tab. \ref{Tab7} we see that all relevant intersections are:
\bea
D_1 \cap D_4 &=&\P^1\,,\qquad D_1 \cap D_5 = \emptyset\,, \qquad\quad\, D_1\cap D_7 = {\cal C}_2\,, \nn \\
D_4 \cap D_5 &=&{\cal C}_2\,,\qquad D_4\cap D_7=2\,\P^1\,,\qquad D_5\cap D_7={\cal C}_2\,. \nn
\eea
Thus we can have winding corrections only from the 17, 45 and 57 intersections since $D_1$ does not intersect with $D_5$ and a $\P^1$ does not contain non-contractible 1-cycles. The volumes of the relevant intersection curves read:
\be
t^\cap(D_1 \cap D_7) =  t^\cap(D_4 \cap D_5) = t^\cap(D_5 \cap D_7) = 6 \, t_f\,.
\ee
Therefore from (\ref{VgsW}) we have that winding string loop corrections take the form:
\be
V_{g_s}^\W=-2\left(\frac{g_s}{8\pi}\right) \frac{W_0^2}{\vo^3} \left( \frac{C_{\rm tot}^\W}{6 \, t_f}\right)
= -\left(\frac{g_s}{8\pi}\right) \frac{W_0^2}{\vo^3} \frac{C_{\rm tot}^\W}{\sqrt{\tau_f}}\,\quad\text{with}\quad C_{\rm tot}^\W=C_1^\W+C_2^\W+C_3^\W\,.
\label{VWnew}
\ee
Given that the D7s on $D_1$ do not intersect with the D7s on $D_5$, there are also KK loop corrections arising from parallel D7 stacks. Since the volume of $D_5$ is given by $\tau_f$ and from (\ref{Ds}) $D_1 = D_5 - D_2$ with $D_2$ a blow-up mode with small volume, the simple expression for the volume (\ref{SimpVol}) suggests that the distance between the two D7-stacks is parametrised by the base of the fibration $t_b$. Hence, using (\ref{LoopCorrKkk}), this KK string loop correction to the K\"ahler potential becomes:
\be
K^\KK_{g_s, {\rm D7/D7}} = g_s \frac{C_b^\KK t_b}{\vo} \,.
\label{77corr}
\ee
Given that the fixed point set shown in Tab. \ref{Tab6} includes also 2 O3-planes located at $D_1 D_2 D_3$ on the CY hypersurface, additional KK $g_s$ effects arise from O7/O3 and D7/O3 combinations. Including the D7/D7 loop correction (\ref{77corr}), the total 1-loop KK scalar potential is given by (\ref{VgsKK}) with a sum over all basis elements. Ignoring terms which depend only on $\vo$ and $\tau_s$ that are fixed at leading order, and neglecting terms which have the same volume scaling as (\ref{VWnew}) but with additional suppression powers of $g_s^2\ll 1$, we end up with focusing on the region where $\sqrt{\tau_f}\tau_b\gg \tau_s^{3/2}$:
\be
V_{g_s}^\KK = g_s^2 \left(\frac{g_s}{8\pi}\right) \frac{W_0^2}{\vo^2} \left[\frac{(C_f^\KK)^2}{4\tau_f^2} + \frac{(C_b^\KK)^2 \tau_f}{72 \vo^2} \left(1-6\,\frac{C_s^\KK}{C_b^\KK} \sqrt{\frac{\tau_s}{\tau_f}}+ 2\,\frac{C_f^\KK}{C_b^\KK} \left(\frac{\tau_s}{\tau_f}\right)^{3/2}\right) \right]\,.
\label{VKKnew}
\ee
Therefore the sum of the two string loop corrections (\ref{VWnew}) and (\ref{VKKnew}) to the scalar potential for the brane setup 1 is:
\be
V_{g_s}  = \left[\frac{A}{\tau_f^2}-\frac{B}{\vo\sqrt{\tau_f}}+\frac{\tau_f}{\vo^2}
\left(C- \tilde{C} \sqrt{\frac{\tau_s}{\tau_f}}+ \hat{C} \left(\frac{\tau_s}{\tau_f}\right)^{3/2}\right)\right] \frac{W_0^2}{\vo^2}\,,
\label{VA1new}
\ee
where:
\bea
A &=& \frac{\kappa}{4} \left(g_s\,C_f^\KK\right)^2 >0 \nn \\
B &=&  \kappa\, C_{\rm tot}^\W \nn \\
C &=& \frac{\kappa}{72} \left(g_s\, C_b^\KK\right)^2 >0 \nn \\
\tilde{C} &=&  \frac{\kappa}{12}\, g_s^2\,C_s^\KK\,C_b^\KK  \nn \\
\hat{C} &=&   \frac{\kappa}{36}\,g_s^2\,C_f^\KK\,C_b^\KK\,.  \nn
\eea
Notice that in the field region where $\tau_f\gg \tau_s$ the terms in (\ref{VA1new}) proportional to $\tilde{C}$ and $\hat{C}$ are negligible and the loop-generated scalar potential simplifies to:
\be
V_{g_s}  \simeq \left(\frac{A}{\tau_f^2}-\frac{B}{\vo\sqrt{\tau_f}} +\frac{C\,\tau_f}{\vo^2} \right) \frac{W_0^2}{\vo^2}\,.
\label{VA1simplnew}
\ee
This reproduces exactly the inflationary potential of `fibre inflation' \cite{Cicoli:2008gp}.

\subsubsection{Case 2}

In this case D7-tadpole cancellation is ensured by placing 4 D7-branes (plus their images) on top of each of two O7-planes which wraps respectively $D_1$ and $D_5$. From Tab. \ref{Tab7} we see that $D_1 \cap D_5 = \emptyset$, and so there is no intersection between the O7s and the D7s, resulting in the absence of winding loop corrections. 

KK loop effects can instead arise from either non-intersecting O7/D7 systems or from O7/O3 and D7/O3 combinations. In the first case, KK string loop correction to the K\"ahler potential take the same for as in (\ref{77corr}). In the second case (the fixed point set has 2 O3-planes at $D_2 D_3 D_4$), since the volume of $D_5$ is given by $\tau_f$, the simple expression for the volume (\ref{SimpVol}) suggests that the distance between the O7s/D7s and the O3s is parametrised again by the base of the fibration $t_b$. Hence this second type of KK loop corrections to the K\"ahler potential take again the same form as in (\ref{77corr}). Using (\ref{VgsKK}), the final KK string loop correction to the scalar potential become:
\be
V_{g_s}^\KK = \frac{C\,\tau_f\,W_0^2}{\vo^4} \qquad\text{with}\qquad C = \frac{\kappa}{72}\left(g_s\,C_b^\KK\right)^2 \,.
\label{VKKstring}
\ee

\subsubsection{Case 3}

This brane setup is characterised by 3 D7-stacks wrapping $D_1$, $D_5$ and $D_7$ and an O7-plane located at $D_6$. From Tab. \ref{Tab7} we see that all relevant intersections are:
\bea
D_1 \cap D_6 &=&\mc{C}_4\,,\qquad \,\,\,D_1 \cap D_5 = \emptyset\,, \qquad\quad\, D_1\cap D_7 = {\cal C}_2\,, \nn \\
D_5 \cap D_6 &=&\mc{C}_{10}\,,\qquad D_6\cap D_7=6\,\P^1\,,\qquad D_5\cap D_7={\cal C}_2\,. 
\label{Inter}
\eea
Thus we can have winding corrections only from the 16, 17, 56 and 57 intersections since $D_1$ does not intersect with $D_5$ and a $\P^1$ does not contain non-contractible 1-cycles. The volumes of the relevant intersection curves read:
\be
t^\cap(D_5 \cap D_6)= 3\,t^\cap(D_1 \cap D_7)  = 3\,t^\cap(D_5 \cap D_7) = 18 \, t_f\,,\quad t^\cap(D_1 \cap D_6) = 4\,t_s+18\,t_f\,.
\ee
Thus using (\ref{VgsW}) we find that winding string loop corrections look like:
\be
V_{g_s}^\W=-2\left(\frac{g_s}{8\pi}\right) \frac{W_0^2}{\vo^3} \left( \frac{C_1^\W}{6 \, t_f}+ \frac{C_2^\W}{6 \, t_f+\frac43\,t_s} \right)
= -\left(\frac{g_s}{8\pi}\right) \frac{W_0^2}{\vo^3} \left(\frac{C_1^\W}{\sqrt{\tau_f}}+\frac{C_2^\W}{\sqrt{\tau_f}-\frac23\sqrt{\tau_s}}\right)\,.
\label{VWnew3}
\ee
Tab. \ref{Tab6} shows that this orientifold involution does not give rise to any O3-plane. Thus KK loop correction can arise only from non-intersecting D7/O7 pairs. From (\ref{Inter}), we see that only the D7s on $D_1$ do not intersect the D7s on $D_5$. Hence the only KK loop correction to the K\"ahler potential takes the form given in (\ref{77corr}) which at the level of the scalar potential yields the correction given in (\ref{VKKstring}).

\subsection{Higher derivative corrections}

Higher derivative $\alpha'^3$ corrections to the scalar potential behave as in (\ref{VF4}). The topological quantities $\Pi_i$'s in our case take become: 
\be
\Pi_1 =140\,,\quad \Pi_2 =44\,,\quad \Pi_3 =44\,, \quad \Pi_4 =8\,, \quad \Pi_5 =24\,,\quad \Pi_6 = 16\,,  \quad \Pi_7 =4\,,\quad \Pi_x = 108\,. \nn
\ee
Focusing on $\Pi_2$, $\Pi_5$ and $\Pi_x$ we find the following scalar potential contribution:
\be
V_{F^4} = - \left(\frac{g_s}{8\pi}\right)^2 \frac{\lambda\, W_0^4}{g_s^{3/2}\vo^4} \left(44 \,t_s + 24 \,t_b + 108\, t_f\right)
\simeq \left(\frac{F}{\vo\,\tau_f}+\frac{G \, \sqrt{\tau_f}}{\vo^2} \right) \frac{W_0^2}{\vo^2}\,,
\ee
where we have neglected the term independent on the fibre modulus $\tau_f$ and: 
\be
F = -24 \,\lambda \left(\kappa\, W_0\right)^2 g_s^{-3/2} \qquad\text{and}\qquad G = \frac32\,F\,.
\ee

\end{document}